\documentclass[prb,twocolumn,superscriptaddress,longbibliography]{revtex4-2}

\usepackage{amsmath,amsfonts,amsthm,bm,amssymb} 
\usepackage{graphicx}
\usepackage{braket}
\usepackage{physics}
\usepackage{dcolumn}
\usepackage{dsfont}
\usepackage[dvipsnames]{xcolor}
\usepackage{gensymb}
\usepackage{soul}

\usepackage[unicode=true,
 bookmarks=true,bookmarksnumbered=true,bookmarksopen=false,
 breaklinks=true,pdfborder={0 0 0},backref=false,colorlinks=true]
 {hyperref}

\hypersetup{
citecolor={blue},
urlcolor={blue}
}

\newcommand{\trs}{\mathcal{T}}
\renewcommand{\emph}{\textit}

\begin{document}
\title{Particle-hole asymmetric phases in doped twisted bilayer graphene}
\author{Run Hou}
\affiliation{Department of Physics and Astronomy, Rice University, Houston, Texas 77005, USA}
\author{Shouvik Sur}
\affiliation{Department of Physics and Astronomy, Rice University, Houston, Texas 77005, USA}
\author{Lucas K. Wagner}
\affiliation{Department of Physics, University of Illinois, Urbana-Champaign, Illinois 61801, USA}
\author{Andriy H. Nevidomskyy}
\affiliation{Department of Physics and Astronomy, Rice University, Houston, Texas 77005, USA}
\affiliation{Rice Center for Quantum Materials and Advanced Materials Institute, Rice University, Houston, Texas 77005, USA}

\begin{abstract}
Twisted bilayer graphene (TBG) has emerged as a paradigmatic platform for exploring the interplay between strong interactions in a multiband system with nearly flat bands, while offering unprecedented control over the filling fraction of electron and hole carriers.
Despite much theoretical work, developing a comprehensive \textit{ab initio} model for this system has proven challenging due to the inherent tradeoff between accurately describing the band structure and incorporating the interactions within the Hamiltonian, particularly given the topological obstruction -- so-called fragile topology -- to the description of the model in terms of localized symmetric Wannier functions within the flat band manifold.
Here, we circumvent this obstruction by using an extended eight-orbital model, for which localized Wannier orbitals have been formulated by [\href{https://doi.org/10.1103/PhysRevResearch.1.033072}{Carr \emph{et al.}, Phys. Rev. Res. 1, 033072 (2019)}]. We constructed an extended multiorbital Hubbard model, and performed Hartree-Fock (HF) calculations to explore its phase diagram across commensurate fillings from -3 to 3. 
We found several nearly degenerate insulating states at charge neutrality, all of which exhibit orbital orders. Crucially, TBG near magic angle is known to be particle-hole asymmetric, which is naturally captured by the single-particle band structure of our model and is reflected in the distinction between the symmetry broken states obtained at electron and hole dopings away from the charge neutral point.
At filling $-1$ and +2, quantum anomalous hall state and inter-orbital intervalley coherent states are obtained, while  
for the rest of the integer fillings away from charge neutrality, we found the system to realize metallic states with various orbital, valley and spin orderings. 
We also observed that most of the Hartree--Fock ground states exhibit a generalized valley Hund's-like rule, resulting in valley polarization. 
Importantly, we show that the incorporation of the intravalley and intervalley exchange interactions is crucial to properly stabilize the ordered symmetry-broken states.
In agreement with experiments, we find significant particle-hole asymmetry, which underscores the importance of using particle-hole asymmetric models.
\end{abstract}

\maketitle

\section{Introduction}\label{sec:intro}
The theoretical prediction of nearly flat bands~\cite{bistritzer2011moire} and subsequent discovery of correlated insulating and superconducting phenomena~\cite{cao_correlated_2018,cao_unconventional_2018} in magic-angle twisted bilayer graphene (TBG) has sparked significant interest in the interplay between topology and strong correlations, leading to extensive theoretical~\cite{moon2012energy,kang_symmetry_2018,koshino_maximally_2018,zou2018,po_fragile_2018,Efimkin2018TBG,isobe2018unconventional,liu2018chiral,xu2018topological,ochi_possible_2018,xux2018,po_origin_2018,guinea2018,guo2018pairing, Wu2018TBG-BCS,yuan2018model,venderbos2018,padhi2018doped,wux2018b,thomson2018triangular,kennes2018strong,dodaro2018phases,song_all_2019,hejazi_multiple_2019,tarnopolsky_origin_2019,daliao_VBO_2019,liu2019pseudo,hejazi_landau_2019,you2019,Lian2019TBG,classen2019competing,zhang2019nearly,kang_strong_2019,seo_ferro_2019,hu2019_superfluid,pixley2019,huang2019antiferromagnetically,liu2019quantum,gonzalez2019kohn,lian2020,padhi2020transport,wu_collective_2020,bultinck2020,bultinck_ground_2020,hejazi2020hybrid,khalaf_charged_2020,xie_superfluid_2020,julku_superfluid_2020,kang_nonabelian_2020,Chichinadze2020Nematic,soejima2020efficient,knig2020spin,christos2020superconductivity,lewandowski2020pairing,Kwan2020Twisted,Parameswaran2021Exciton,xie_HF_2020,liu2020theories,cea_band_2020,zhang_HF_2020,liu2020nematic,daliao2020correlation,eugenio2020dmrg,huang2020deconstructing,Calder2020Interactions,ledwith2020,repellin_EDDMRG_2020,abouelkomsan2020,repellin_FCI_2020,vafek2020hidden,fernandes_nematic_2020,Wilson2020TBG,wang2020chiral,TBG1,Song-TBG2,Biao-TBG3,Biao-TBG4,TBG5,TBG6,zhang2021momentum,vafek2021lattice,cha2021strange,sheffer2021chiral,kang2021cascades,hofmann2021fermionic,Thomson2021Recovery,kwan2021kekule,parker2021strain,Pathak2022,wagner2022global,song2022magic,Shi_heavy_fermion2022,borovkov2023tbg,xie2023phase,hukondo2023,husymmetric2023,choukondo2023,2023arXiv230113024D,su4itierant} and experimental studies~\cite{lu2019superconductors, yankowitz2019tuning, sharpe_emergent_2019, serlin_QAH_2019, polshyn_linear_2019,  xie2019spectroscopic, choi_imaging_2019, kerelsky_2019_stm, jiang_charge_2019,saito_independent_2020, stepanov_interplay_2020, liu2021tuning, arora_2020, cao_strange_2020,wong_cascade_2020, zondiner_cascade_2020,  nuckolls_chern_2020, choi2020tracing, saito2020,das2020symmetry, wu_chern_2020,park2020flavour,saito2020isospin,rozen2020entropic,stepanov_competing_2021,lu2020fingerprints,Grover2022,Jaoui2022,das2022observation} in the field of condensed matter physics. The construction of effective models and the development of theories to understand this system have emerged as crucial endeavors in contemporary research.

In order to establish models that are both simple and capable of capturing the essential features of magic-angle TBG without sacrificing the \emph{ab initio} perspective, various approaches have been explored. These include models based on  maximally localized Wannier functions (MLWF)~\cite{kang_symmetry_2018,koshino_maximally_2018,carr2019derivation}, atomic tight-binding~\cite{moon2012energy,Pathak2022}, the Bistritzer--MacDonald (BM) model~\cite{bistritzer2011moire}, and topological heavy fermion (THF) model~\cite{Shi_heavy_fermion2022,song2022magic,borovkov2023tbg,choukondo2023}. However, developing a comprehensive model has proven challenging due to the inherent tradeoff between accurately describing the band structure and incorporating the interactions within the Hamiltonian, particularly given the fragile topological nature~\cite{zou2018,po_fragile_2018} of the magic-angle TBG system.

The BM model, despite its simplicity in constructing the noninteracting band structure of magic-angle TBG, is approximate due to its inherent \textbf{k}$\cdot$\textbf{p} nature and moreover, encounters complexity when incorporating interactions by projecting out remote high-energy bands. This renders the traditional methods for studying strongly correlated systems less applicable. Alternatively, the THF model for TBG has been developed and has yielded promising results~\cite{hukondo2023,husymmetric2023,choukondo2023}. However, MLWF models offer another viable option as they provide accurate descriptions of the band structure and localized interactions, albeit with a higher number of orbitals. Regrettably, the MLWF models as alternative choices for magic-angle TBG have not been extensively investigated.

Therefore, in this study, we adopt an eight-orbital model of magic-angle TBG from Carr~\textit{et al.}~\cite{carr2019derivation} and employ Hartree--Fock (HF) methods to explore its properties. 
In order to minimize an \textit{a priori} bias in the symmetry-broken states thus obtained, we do not constrain the density matrices in either the spin or valley channels to be identical, and we allow mixing between different species of electrons.
Unlike the belief that Wannierized models are challenging to use in calculations, we found the HF method to be straightforward, as it does not involve complicated form factors such as those arising from projecting the interactions onto the BM model.
The 8-orbital model also has the potential for future use in more sophisticated strongly correlated methods such as DMFT, to calculate the states that cannot be captured by the HF self-consistent calculations. Although previous studies~\cite{Calder2020Interactions} had considered the 8-band model, those calculations were done at the Hartree level (without the Fock terms), and did not find any interaction-driven insulating states.

In this work, we first used the Wannier orbitals from the 8-orbital model to numerically evaluate the strength of the electron repulsion integral matrix elements, enabling us to construct a suitably extended Hubbard model including both the direct and exchange interactions. Subsequently, we incorporated the complete Hartree and Fock terms to study this model. 
We identified several converged ordered states for each filling. These symmetry-broken ordered states originate from the valley, spin, and orbital degrees of freedom. Our results compare favourably with the experimental findings~\cite{yankowitz2019tuning,sharpe_emergent_2019,serlin_QAH_2019,polshyn_linear_2019,jiang_charge_2019,stepanov_interplay_2020,liu2021tuning,saito2020,park2020flavour,lu2019superconductors,saito_independent_2020,Jaoui2022,das2022observation}, including the quantum anomalous Hall (QAH) states found at certain fillings. The particle-hole asymmetry is naturally present in our 8-orbital model, in contrast to the BM model, and is consistent with the experimentally observed asymmetry between electron- and hole-doped side of the phase diagram.
A detailed comparison between our results and experimental findings suggests that the symmetry-breaking and gap formation mechanisms depend on the filling fraction and the specific device configurations used in the experiments. Our HF calculations underscore the importance of incorporating particle-hole asymmetric models into theoretical considerations.

We briefly discuss the mechanisms of spontaneous symmetry breaking here. 
Some previous studies had used (approximate) valley-spin SU(4) symmetry to discuss various symmetry-broken states in TBG systems~\cite{xu2018topological,su4itierant}. The atomistic model, such as the one employed here~\cite{carr2019derivation}, does not actually possess full SU(4) symmetry -- instead the kinetic energy terms have $\text{U(2)}_+ \times \text{U(2)}_-$ symmetry where $\pm$ label the two graphene valleys, with the $\text{U(2)}=\text{U(1)}_\text{charge}\times \text{SU(2)}_\text{spin}$ symmetry separately in each valley.
Nevertheless, there are close parallels with the spontaneously broken valley and spin symmetries that are also present in SU(4)-type models~\cite{su4itierant}, or in the generalized Stoner picture~\cite{saito_independent_2020}. We summarize the various possible spontaneous symmetry-broken states in Table~\ref{table:stoner}.

Importantly,  we found that the atomistic model also contains information about orbital degrees of freedom ($p_+$ and $p_-$ orbitals in the 8-band model~\cite{carr2019derivation}), which is absent from the BM-type analysis and from the above symmetry considerations. These orbitals are energetically degenerate in the non-interacting limit, and this additional symmetry can be spontaneously broken in the presence of interactions. Indeed, we found this to be the case in nearly all of the ordered states we identified, with orbital degeneracy spontaneously lifted. 
In fact, we find the ground state to be a combination of orbital ordering and one of the symmetry-broken patterns summarized in Table~\ref{table:stoner}, depending on the filling fraction.

\begin{table*}[ht]
\centering
\begin{tabular}{ c c }
\hline
\hline
States & Symmetry  \\
\hline
Normal state & $\text{U(2)}_+ \times \text{U(2)}_-$\\
(a)Valley polarized (VP) state& Breaking time-reversal symmetry $\trs$\\
(b)Orbital polarized (OP) state& Breaking discretized symmetries such as $C_{2z}\trs$ and $C_{2x}$\\
(c)Spin polarized (SP) state& $\text{U(1)}_+ \times \text{U(2)}_-$ or  $\text{U(1)}_+ \times \text{U(1)}_-$\\
(d)Inter-valley coherent (IVC) state &$\text{SU(2)}_+ \times \text{SU(2)}_- \times \text{U(1)}$\\
\hline
\hline
\end{tabular}
\caption{Examples of simplified symmetry-broken orders. When the interaction strength is large enough, the system develops orders. Realistic models produce more complicated orders. (a) The VP state breaks time-reversal symmetry. (b) The OP state breaks the orbital discretized symmetry. (c) The SP state breaks the spin SU(2) symmetry. (d) The IVC state breaks the $\text{U(1)}_+ \times \text{U(1)}_-$ symmetry.} 
\label{table:stoner}
\end{table*}

This paper is organized as follows. In Sec. \ref{sec:hamiltonian}, we review the non-interacting 8-orbital tight-binding model from previous work and numerically evaluate the interaction parameters. In Sec. \ref{sec:N}, we discuss the numerical results, where we present the order parameters, symmetry breaking, and associated Chern numbers, if relevant, of the obtained HF solutions. Importantly, we also show that the role of exchange interactions is crucial to correctly capture the correlated insulating states of  the TBG. We compare our numerical results with other typical HF calculations in Sec. \ref{ref:HFComparision}. We compare our findings to the experiments in Sec.~\ref{sec:comparison}, before summarizing our conclusions in Sec.~\ref{sec:conclusion}.

\section{MINIMAL INTERACTING TIGHT-BINDING MODEL FOR MATBLG}\label{sec:hamiltonian}
In this section, we review the lattice structure and construct the interacting 8-orbital model of magic-angle TBG.

\subsection{Lattice structure and noninteracting Hamiltonian}
\begin{figure}[!htb]
    \centering
    \includegraphics[width=\linewidth]{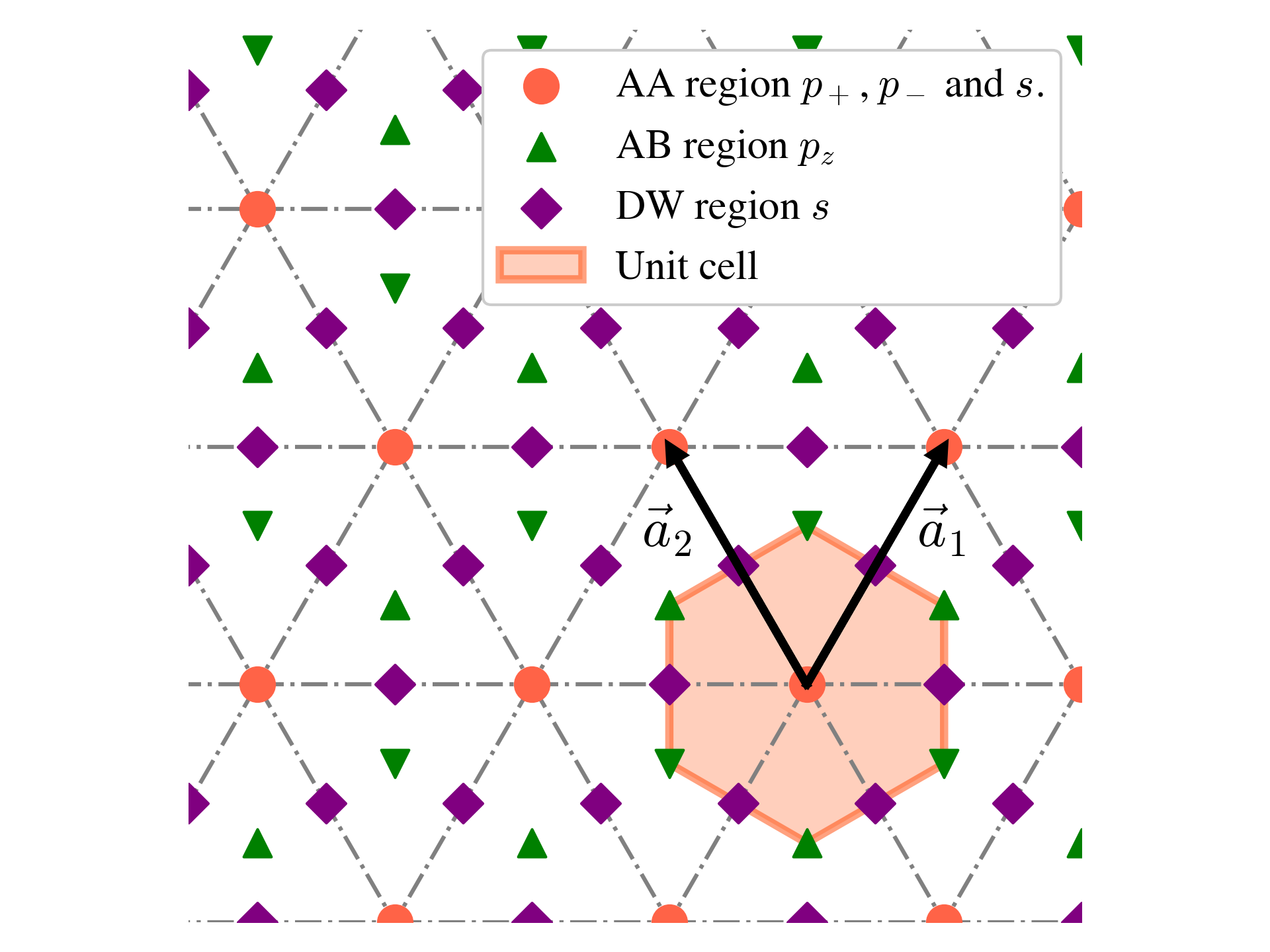}
    \caption{Lattice structure of the 8-orbital model. The Bravais lattice of the system is a triangular lattice. Orange dots in triangular lattice represent orbital 1/2 with $p_{\pm}$ symmetries and orbital 3 with $s$ symmetry. Green triangles in honeycomb lattice represent $p_z$ orbitals labeled by orbital indices 4 and 5. Purple diamonds in the kagome lattice represent orbitals 6,7, and 8 with $s$ symmetry. The primitive vectors are $\vec{a}_1$ and $\vec{a}_2$.}
    \label{fig:lattice}
\end{figure}

To begin with, we review the lattice structure of TBG eight-band model derived in \cite{carr2019derivation}. There are eight Wannier orbitals per spin and per valley inside a given triangular unit cell of the moir\'{e} lattice as shown in Fig.~\ref{fig:lattice}. The orbital indices, corresponding Wannier orbital centers, and orbital symmetries are shown in TABLE~\ref{table1}. 
This eight-orbital tight binding model has correct crystalline symmetries, avoiding the Wannier obstruction resulting from fragile topology if only a subset of bands are included~\cite{po_fragile_2018}.

\begin{table}[ht]
\centering
\begin{tabular}{ c c c c c }
\hline
\hline
Orbital indices  & Region & Wyckoff position& Symmetry \\
\hline
 1,2& AA&$1a$ & $p_+,p_-$\\
 
 3 &AA&$1a$&$s$\\

 4,5 &AB&$2b$ &$p_z$\\

 6,7,8&DW&$3c$&$s$\\
 \hline
 \hline
\end{tabular}
\caption{Orbital indices and corresponding regions and symmetries. The AA region is located at crossing of triangular lattice usually denoted as $1a$. The AB region is located at the center of triangles denoted as $2b$. And the DW region is located at the centers of triangular edges denoted as $3c$.  }
\label{table1}
\end{table}

The Wannier orbitals and noninteracting Hamiltonian is obtained from the maximally localized Wannier functions (MLWF) method, and the resulting noninteracting tight-binding model is taken from Carr \textit{et al.}~\cite{carr2019derivation}:
\begin{equation}
    H_K = \sum_{ij,ab,\tau s} t^{\tau}_{ab}(\bm{R}_i-\bm{R}_j) c^\dagger_{a \tau s}(\bm{R}_i)c_{ b\tau s}(\bm{R}_j),
\end{equation}
where $c^\dagger_{a \tau s}(\bm{R}_i)$ is the electron creation operator in the $a^\text{th}$ Wannier orbital ($a=1,\ldots 8$) positioned within the moir\'e unit cell labeled with a lattice vector $\bm{R}_i$. The subscript $\tau=\pm$ denotes graphene valley indices, and $s=\uparrow,\downarrow$ are the electron spin indices. 
The  hopping parameters $t^\tau_{ab}(\bm{R_i}-\bm{R_j})$ are fitted from \emph{ab initio} $\bm{k} \cdot \bm{p}$ model with lattice relaxation. In the present work, we choose as a basis the noninteracting Hamiltonian at twist angle $\theta = 1.10\degree$.

We Fourier transform the electron creation/annihilation operators into momentum space using the periodic gauge, meaning that the Fourier phases do not depend on specific positions of Wannier centers but rather only on the moir\'e cell coordinate  $\bm{R_i}$: 
\begin{equation}
    c^\dagger_{a \tau s}(\bm{R}_i) = \frac{1}{\sqrt{N_k}}\sum_{\bm{k}}c^\dagger_{a \tau s}(\bm{k}) e^{i\bm{R}_i\cdot\bm{k}}.
\end{equation}

The momentum-space noninteracting Hamiltonian is then written in the form
\begin{equation}
    H_K = \sum_{ab\tau s} t^{\tau}_{ab}(\bm{k}) c^\dagger_{a \tau s}(\bm{k})c_{b\tau s}(\bm{k}).
\end{equation}
The periodic gauge has the convenient property that $t^{\tau}_{ab}(\bm{k}+\bm{G}) = t^{\tau}_{ab}(\bm{k})$, where $\bm{G}$ is the moir\'e reciprocal lattice vector. 

The noninteracting Hamiltonian has several important symmetries. The time-reversal symmetry relates the hoppings in the two valleys via
\begin{equation}
    t^{-\tau}_{ab}(\bm{k}) = t^{\tau*}_{ab}(-\bm{k}).
\end{equation}
The Hamiltonian is also invariant under operations of $C_{2z}\trs$, $C_{3z}$, and $C_{2x}$ symmetries as demonstrated in Appendix~\ref{app: symm}. For the internal symmetries, it is clear that the system preserves $\text{U(2)}_+\times \text{U(2)}_-$ symmetry, which means for each valley $+/-$ the system has spin SU(2) symmetry and preserves the total particle number $N_+ + N_-$ and the particle number difference $N_+ - N_-$.

\begin{figure*}
    \centering
    \includegraphics[width=\linewidth]{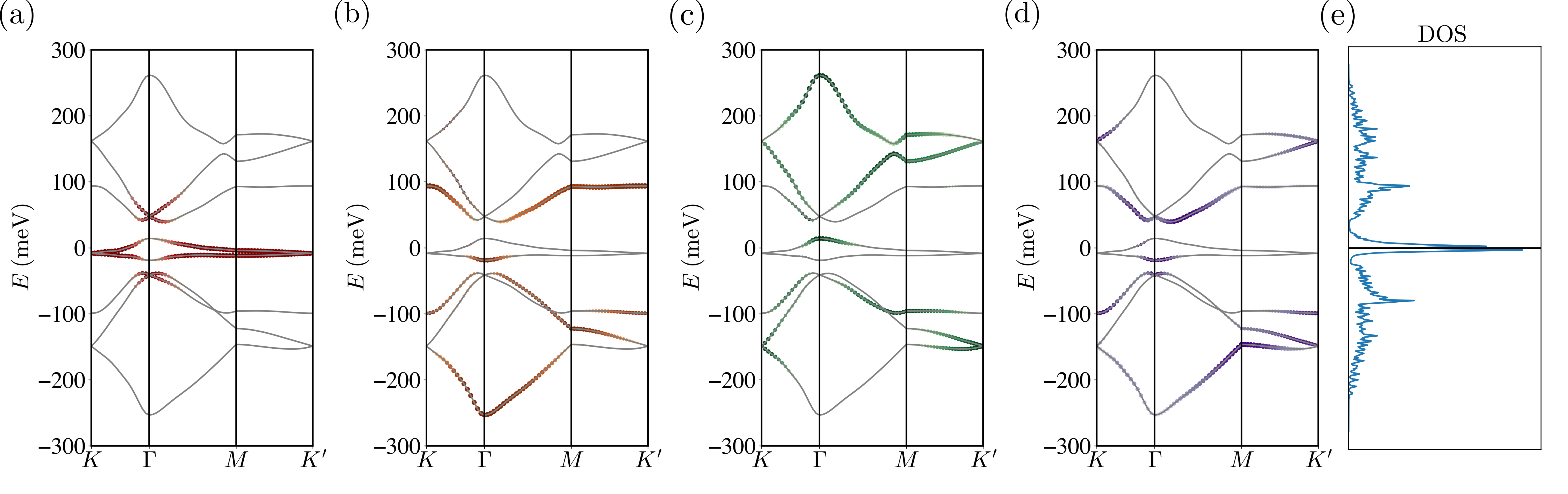}
    \caption{The tight-binding band structure of non-interacting Hamiltonian. (a) The dark orange fat bands show the projected weights of $AA_{p_{\pm}}$ orbitals. (b) The orange fat bands show the projected weights of $AA_s$ orbital. (c) The fat bands for $AB_{p_z}$ orbitals. (d) The fat bands for $DW_s$ orbitals. (e) The density of states of the whole band structure. }
    \label{fig:band}
\end{figure*}

In Fig.~\ref{fig:band}, we showed the non-interacting band structure of the single valley 8-orbital TBG model along with the projected weights of orbitals that have different orbital symmetries. It clearly shows that the central bands are mainly composed by $p_\pm$ orbitals except for the points near $\Gamma$ point.

\subsection{Interacting Hamiltonian}

The full Hamiltonian is composed of the noninteracting Hamiltonian $H_K$ and interactions $H_I$:
\begin{equation}
    H = H_K + H_I.
\end{equation}

Now, we elaborate how to define the interactions in our multiorbital model.
The most general form of the interacting Hamiltonian that preserves spin SU(2) symmetry (since the spin-orbit coupling is negligible in graphene) is as follows:
\begin{multline}
    H_{I}=\frac{1}{2}\sum_{\tau_j, a_j, \mathbf{R}_j, s,s^\prime}
    V^{\tau_1\tau_2\tau_3\tau_4}_{a_1 a_2 a_3 a_4}(\bm{R}_1,\bm{R}_2,\bm{R}_3,\bm{R}_4)\times\\
    c^\dagger_{a_1 \tau_1 s}(\bm{R}_1)c^\dagger_{a_2\tau_2 s^\prime}(\bm{R}_2)
    c_{a_3\tau_3 s^\prime}(\bm{R}_3)c_{a_4\tau_4 s}(\bm{R}_4).
\end{multline}
The four-center electron repulsion integrals $V^{\tau_1\tau_2\tau_3\tau_4}_{a_1 a_2 a_3 a_4}(\bm{R}_1,\bm{R}_2,\bm{R}_3,\bm{R}_4)$ were evaluated using the real-space Wannier functions, and we found the oscillating factor $e^{i\bm{q}_V \cdot (\bm{r}-\bm{r}^\prime)}$
suppresses the absolute value of the integral when $\tau_2 \neq \tau_3$ and $\tau_1 \neq \tau_4$, because the momentum $\bm{q}_V$ between the graphene two valleys is large as explained in Appendix~\ref{app: interaction}. With the benefit of MLWF, it is reasonable to only evaluate the 2-center interactions that have the forms like  $V^{\tau\tau^\prime \tau^\prime \tau}_{a_1 a_2 a_3 a_4}(\bm{R}_1,\bm{R}_2,\bm{R}_2,\bm{R}_1)$ and $V^{\tau\tau^\prime \tau^\prime \tau}_{a_1 a_2 a_3 a_4}(\bm{R}_1,\bm{R}_2,\bm{R}_1,\bm{R}_2)$.

Among these interaction channels, we found the strongest interaction channels are density-density channels $H_C$, with the exchange channels $H_X$ playing a key role away from half-filling.
This motivates the form of the Hamiltonian
\begin{equation}
    H = H_K + H_C +H_X.
\label{eq:Hfull} 
\end{equation}
The density-density interacting Hamiltonian takes the form
\begin{equation}
\begin{split}
    H_{C}=\frac{1}{2}\sum_{ij;ab;\tau\tau^\prime s s^\prime}
    U_{ab}(\bm{R}_i-\bm{R}_j)\times\\
    c^\dagger_{a\tau s}(\bm{R}_j)c^\dagger_{b\tau^\prime s^\prime}(\bm{R}_i)
    c_{b\tau^\prime s^\prime}(\bm{R}_i)c_{a\tau s}(\bm{R}_j),
\end{split}
\label{eq:direct}
\end{equation}
which is an extended Hubbard model with long-range interaction. The density-density repulsive interaction $U_{ab}(\bm{R}_i-\bm{R}_j)$ does not depend on the choice of valleys and is explicitly evaluated assuming a screened single-gated Coulomb potential $V_{SG}(r)$ often used in modeling of TBG:
\begin{equation}
    V_{SG}(r) = \frac{1}{4\pi\epsilon\epsilon_0}\left(\frac{1}{r} - \frac{1}{\sqrt{r^2+(2d)^2}}\right).
\end{equation}
The relative dielectric constant $\epsilon = 12$ was set to capture the screening effect from remote upper and lower bands~\cite{Calder2020Interactions}. 
The distance between the sample and capacitor was set to $d=10\text{nm}$, which is comparable with the magic-angle TBG moir\'e lattice length and is close to realistic experimental settings.

The resulting density-density interaction is
\begin{equation}
\label{eq:Uab}
\begin{split}
        U_{ab}(\bm{R}_i-\bm{R}_j)& = \int \mathrm{d}\bm{r}^2\mathrm{d}{\bm{r}^\prime}^2 V_{SG}(|\bm{r}-\bm{r}^\prime|)\times \\
        &\sum_{Y,Y^\prime}|\mathcal{W}^{Y}_{\bm{R}_ia}(\bm{r} )|^2|\mathcal{W}^{Y^\prime}_{\bm{R}_jb}(\bm{r}^\prime)|^2,
\end{split}
\end{equation}
where $\mathcal{W}^Y_{\bm{R}_{i}a}(\bm{r})$ is the Wannier function for the orbital $a$ at the unit cell located at $\bm{R}_i$ for graphene-sublattice $Y$\cite{carr2019derivation}. 

\begin{figure} [!b]
    \centering
\includegraphics[width=\linewidth]{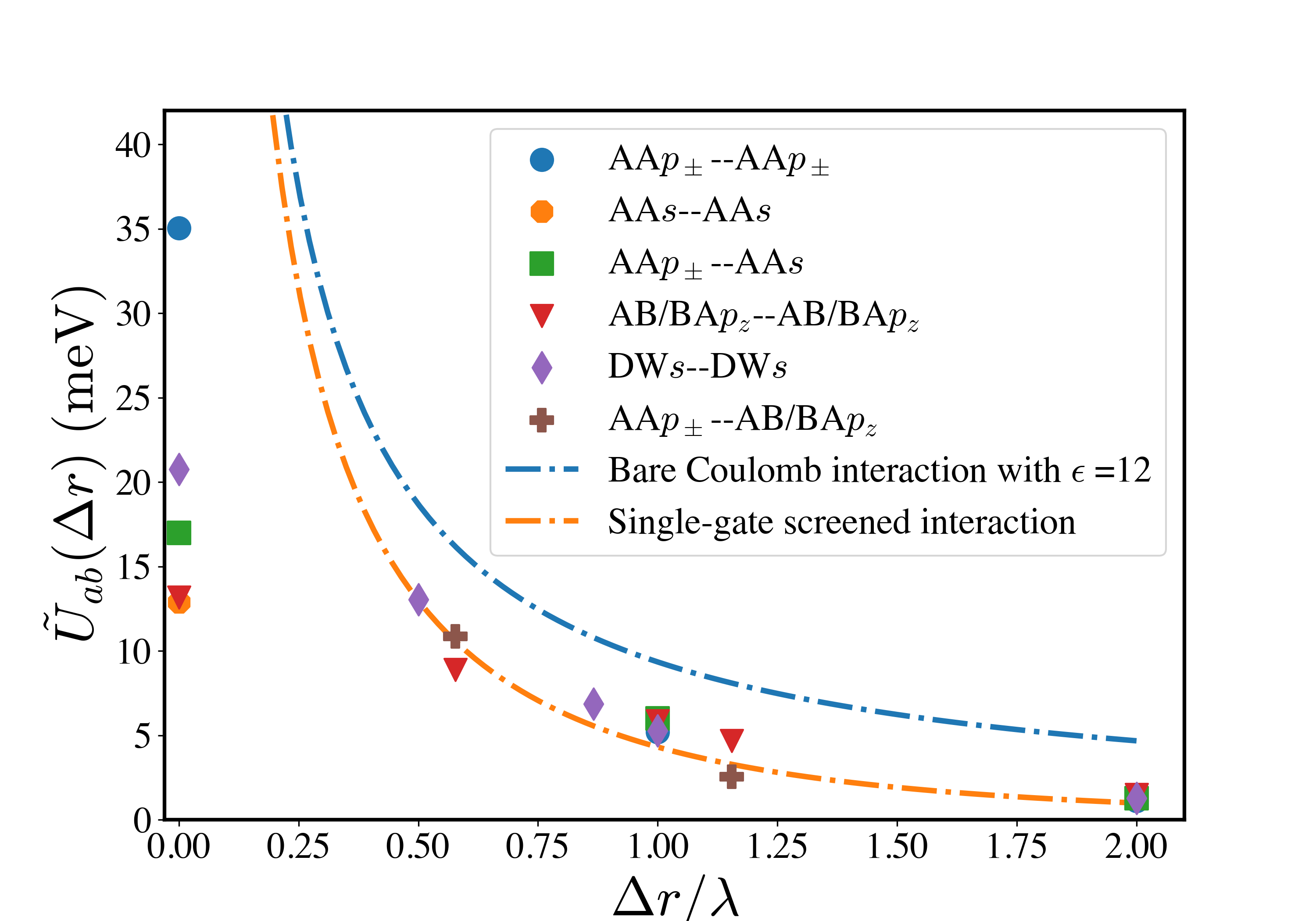}
    \caption{Single-gated screened density-density channel interaction strengths as a function of the distance between orbital centers $\Delta r$. The horizontal axis is rescaled with moir\'e lattice length $\lambda$. Data points with different shapes correspond to different orbital pairs.}
    \label{fig:sg_Coulomb}
\end{figure}

We represent the general trend in the density-density interactions between selected orbitals in Fig.~\ref{fig:sg_Coulomb}.
Noting that the lattice vector $\bm{R}_i$ is not equivalent to the position of the Wannier center $\bm{R}_i + \bm{r}_a$, we elected to plot the interaction integrals  in Fig.~\ref{fig:sg_Coulomb} as a function of the distance $\Delta r = |\bm{R}_i+\bm{r}_a-\bm{R}_j-\bm{r}_b|$ between the two orbital centers in order to compare with the bare single-gated interaction: $\Tilde{U}_{ab}(\Delta r) = U_{ab}(\bm{R}_i-\bm{R}_j)$. The strongest onsite density-density interaction is around 35~meV, which is comparable with the bandwidth of the central bands.
In practice, we chose the density-density repulsive interactions to be extended to next-nearest neighbor unit cells. 
At longer distances, $V_{SG}(r)$ becomes negligible and is smaller than the exchange interactions that we will introduce next.

\begin{figure*} [!t]
    \centering    \includegraphics[width=0.7\linewidth]{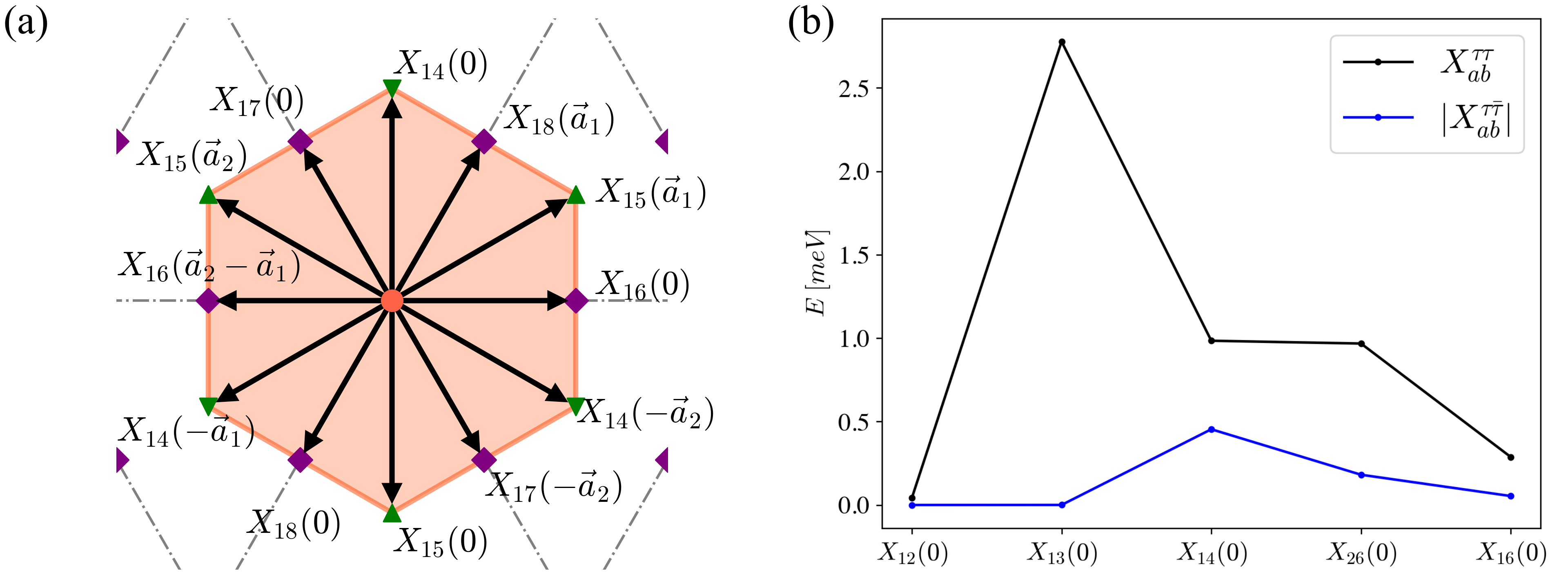}
    \caption{(a) The exchange interaction arrangement. The short-range exchange interactions between $p_{\pm}$ and other orbitals  within a unit cell are taken into account. (b) The exchange interactions in the same valley $X^{\tau \tau}_{ab}(\Delta\bm{R})$ and in different valleys $X^{\tau\bar{\tau}}_{ab}(\Delta\bm{R})$. We found $X^{\tau\bar{\tau}}_{ab}(\Delta\bm{R})$ is negligible in comparison to the exchange interactions in the same valley.}
    \label{fig:sg_exchange}
\end{figure*}

The second part of the interacting Hamiltonian includes the exchange interaction matrix elements $X_{ab}^{\tau\tau'}$ which are obtained as integrals over the corresponding Wannier orbitals, as detailed in Appendix~\ref{app: interaction}:
\begin{equation}\label{eqn:X}
\begin{split}
        &H_X = \frac{1}{2}\sum_{a,b,\tau,\tau^\prime,s, s^\prime}\sum_{ij}X^{\tau\tau^\prime}_{ab}(\bm{R}_i-\bm{R}_j)\times\\
        &:c^\dagger_{a\tau s}(\bm{R}_i)c_{b\tau s}(\bm{R}_j)c^\dagger_{b\tau^\prime s^\prime}(\bm{R}_j)c_{a\tau^\prime s^\prime}(\bm{R}_i):.
\end{split}
\end{equation}

Rather than keep the full complexity of the matrix $X_{ab}^{\tau\tau'}$, we observed that the dominant effect at experimentally accessible densities arises from the $p_{\pm}$ orbitals (which constitute the flat bands in magic-angle TBG) interacting among themselves and with the other orbitals, as illustrated in Fig.~\ref{fig:sg_exchange}(a). In this study, we therefore dropped the matrix elements $X_{ab}^{\tau\tau'}$ whenever both indices $a$ and $b$ $\not\in \{1,2\}$.

Furthermore, while the exchange interaction $X_{ab}^{\tau\tau'}$ depends on the valley indices (see Appendix~\ref{app: interaction} for details), the computed values shown in  Fig.~\ref{fig:sg_exchange}(b) demonstrate that the exchange interaction between different valleys $X^{\tau\Bar{\tau}}_{ab}$, is significantly smaller in magnitude compared to that within the same valley $X^{\tau \tau}_{ab}$. For the sake of simplicity, we assumed $X^{\tau\Bar{\tau}}_{ab} = 0$. 
The largest exchange interaction is approximately $2.7$~meV, corresponding to $X^{\tau \tau}_{13}$ and $X^{\tau \tau}_{23}$, independent of $\tau$,  that is, the exchange between the $p_\pm$ and $s$-orbital centered in the AA region of the same moir\'e supercell. We note that previous studies~\cite{Calder2020Interactions,2023arXiv230113024D} of this model did not consider exchange interactions due to its small magnitude compared  with the on-site density-density interaction. In contrast, in this study we found that exchange interactions have pronounced effects on the interacting model and should not be ignored. In the spirit of \cite{Calder2020Interactions,2023arXiv230113024D} and THF models \cite{song2022magic}, however, we do not consider pair hopping terms.

\section{NUMERICAL RESULTS}\label{sec:N}

\begin{center}
\begin{table*}
    \begin{tabular}{ c  c  c  l  c  c }
\hline
\hline
     $\nu$& Order&Insulator & Order parameter & $C_{2z}\trs$ symmetry& Chern number $|C|$\\
     \hline
     0&OP (orbital-polarized)&\checkmark & $\left<\Sigma_3\right>$&$\cross$&0\\
     0&VOP (valley-orbital polarized)&\checkmark& $\left<\tau_3 \otimes\Sigma_3\right>$&$\cross$&4\\
     0&VOP+OP&\checkmark& $\left<\tau_3 \otimes\Sigma_3\right>_\uparrow, \left<\Sigma_3\right>_\downarrow$&$\cross$&2\\
    0&VP (valley-polarized)&$\cross$& $\left<\tau_3\right>$&\checkmark &N/A\\
    0&SP (spin-polarized)&$\cross$ & $\left<s_3\right> $&$\cross$&N/A\\
    
    0&IVC (inter-valley coherent)&$\cross$ & $\left<\tau_{1,2}\right>$&\checkmark &N/A\\
    0&KIVC (Kramers IVC) &$\checkmark $ & $\left<\tau_{1,2}\otimes \Sigma_3\right>$ & $\cross$&0\\
     0&inter-orbital IVC &$\checkmark $ & $\left<\tau_{1,2}\otimes \Sigma_2\right>$ &$\cross$ &0\\
  
     2&VP&\checkmark&$\left<\tau_3\right>,\left<\Sigma_3\right>,\left<\tau_3\otimes \Sigma_3\right>$&$\cross$&2\\
     2&SP&\checkmark&$\left<s_3\right>,\left<\tau_3\otimes\Sigma_3\right>,\left<s_3\otimes\tau_3\otimes\Sigma_3\right>$&$\cross$&0\\
     2&IVC (inter-valley coherent)+VP&$\cross$&$\left<\tau_3\right>_\uparrow,\left<\tau_{1,2}\right>_\downarrow,\left<\tau_3 \otimes\Sigma_3\right>$&$\cross$&N/A\\

     2&inter-orbital IVC&\checkmark&$\left<s_3\right>,\left<\tau_2\otimes \Sigma_2\right>,\left<s_3 \otimes\tau_2\otimes \Sigma_2\right>$&$\cross$&0\\
   
     -2&VP&$\cross$&$\left<\tau_3\right>,\left<\Sigma_3\right>,\left<\tau_3\otimes \Sigma_3\right>$&$\cross$&2\\
     -2&SP&$\cross$ &$\left<s_3\right>,\left<\tau_3\otimes\Sigma_3\right>,\left<s_3\otimes\tau_3\otimes\Sigma_3\right>$&$\cross$&0\\
     -2&OP&$\cross$&$ \left<\Sigma_3\right>,\left<s_3\otimes\tau_3\right>,\left<s_3\otimes\tau_3\otimes \Sigma_3\right>$&$\cross$&0\\
   
     1&VPSP&$\cross$&$\left<s_3\right>,\left<\tau_3\right>,\left<\Sigma_3\right>$&$\cross$&N/A\\
     1&IVCSP&$\cross$&$\left<s_3\right>,\left<\tau_{1,2}\right>,\left<\Sigma_3\right>$&$\cross$&N/A\\
    
     -1&VPSP&\checkmark&$\left<s_3\right>,\left<\tau_3\right>,\left<\Sigma_3\right>$&$\cross$&1\\
     -1&IVCSP&$\cross$&$\left<s_3\right>,\left<\tau_{1,2}\right>,\left<\Sigma_3\right>$&$\cross$&N/A\\
     
    -3&VPSP&$\cross$&$\left<s_3\right>,\left<\tau_3\right>,\left<\Sigma_3\right>$&$\cross$&N/A\\
     -3&IVCSP&$\cross$&$\left<s_3\right>,\left<\tau_{1,2}\right>,\left<\Sigma_3\right>$&$\cross$&N/A\\
    
     3&VPSP&$\cross$&$\left<s_3\right>,\left<\tau_3\right>,\left<\Sigma_3\right>$&$\cross$&1\\
     \hline
     \hline
\end{tabular}
\caption{Summary of properties of converged states at different filling $\nu$ obtained from HF calculations. The third column labels whether the state is an insulator or not. The fourth column shows the order parameters of the corresponding state. The fifth column shows whether the $C_{2z}\trs$ symmetry is broken. The last column shows the absolute value of the Chern number. We use Pauli matrices $s$, $\tau$, and $\Sigma$ to label spin, valley, and the $p_\pm$ degrees of freedom. The notation $\langle \hat{O}\rangle_{\uparrow/\downarrow}$ represents the order in the spin-up/down species. For IVC states, $\left<\tau_{1,2}\right>$ refers to any order at the $U(1)$ circle formed by $\tau_1$ and $\tau_2$. We calculated the Chern number for all insulating states, also including some states that are not fully gapped -- this is done for reference purposes, if the bands are still separable. The `N/A' denotes the cases where the metallic bands are not separable and the Chern number cannot be computed.}
\label{table:iii}
\end{table*}
\end{center}

\subsection{Order parameters and symmetry breaking}
In this section, we discuss the ground state candidates at the filling fractions $\nu =0, \pm1, \pm2, \pm3$ relevant to the experiment where the correlated insulating behavior has been observed~\cite{cao_correlated_2018,choi_imaging_2019,yankowitz2019tuning,polshyn_linear_2019,xie2019spectroscopic,lu2019superconductors,serlin_QAH_2019,sharpe_emergent_2019,park2020flavour,kerelsky_2019_stm,wu_chern_2020,saito2020,nuckolls_chern_2020,choi2020tracing,saito_independent_2020,stepanov_interplay_2020,das2020symmetry,liu2021tuning,das2022observation}. 
Because the eight-orbital model does not have particle-hole symmetry (unlike the approximate BM model), we found the resulting states not to be the same for the particle-doped and the hole-doped sides of the phase diagram. We performed the multi-band Hartree-Fock (HF) calculations of the Hamiltonian Eq.~\eqref{eq:Hfull} up to $15\cross15$ discretized points in momentum space for a fixed twist angle $\theta = 1.10\degree$. The detailed derivation of the HF theory is presented in Appendix~\ref{app: hftheory}.

 A closer analysis of the converged results shows that all symmetry-broken states can be characterized by spin, valley, and orbital degrees of freedom.
  The most general order parameter is the mean value of fermion bilinears $\Hat{O}(\bm{k}) =\sum_{a\tau s,b\tau^\prime s^\prime}c^\dagger_{a \tau s}(\bm{k})\mathcal{O}_{a\tau s,b\tau^\prime s^\prime}c_{b\tau^\prime s^\prime}(\bm{k}) $:
 \begin{equation}
        \left<\Hat{O}\right> = \frac{1}{N_k}\sum_{\bm{k}}\Tr [\mathcal{O} P(\bm{k})],\label{eq:op}
 \end{equation}
where $P_{a\tau s,b\tau^\prime s^\prime}(\bm{k}) = \left<c^\dagger(\bm{k})_{a \tau s}c(\bm{k})_{b\tau^\prime s^\prime}- \frac{1}{2}\delta_{ab}\delta_{\tau\tau^\prime}\delta_{s s^\prime}\right>$ is the one-body reduced density matrix, and $N_k$ is the total number of momentum points.
While the orbital symmetry breaking is in principle possible for all eight orbitals, we found that all the orbital ordering is limited to the $p_\pm$ orbitals that constitute the nearly flat bands in magic-angle TBG. We thus limited the orbital order to this subset and use the Pauli matrix $\Sigma$ to label the orders formed by the $p_\pm$ orbitals. The ordering of valley and spin degrees of freedom is labeled by the Pauli matrices $\tau$ and $s$, respectively.

While the $C_{2z}\trs$ symmetry is preserved by the tight-binding Hamiltonian, the interactions can result in the symmetry being spontaneously broken. Denoting the  $C_{2z}\trs$ operator by $g$ for brevity, the corresponding symmetry breaking strength
is given by (see Appendix~\ref{app: symmbreaking} for derivation):

\begin{align}
    \mathcal{D}(\bm{k}) = 
    |\left<g^{-1}c^\dagger_{a \tau s}(\bm{k})c_{b\tau^\prime s^\prime}(\bm{k})g\right> - \left<c^\dagger_{a \tau s}(\bm{k})c_{b\tau^\prime s^\prime}(\bm{k})\right>|.
    \label{eq:TRS-op}
\end{align}

When summed over all $\mathbf{k}$-points as in Eq.~\eqref{eq:op}, it measures the difference between the expectation value of the order parameter and its $C_{2z}\trs$ symmetry transformed value. The $C_{2z}\trs$ symmetry-breaking strength is zero if the order parameter is invariant under the $C_{2z}\trs$ symmetry.

Furthermore, we calculated the Chern number $C$ which characterizes quantum anomalous Hall states in insulators. The Chern number is calculated by measuring the winding of the wavefunction along the Wilson loop as shown in Appendix~\ref{app: wilson}.

Table~\ref{table:iii} summaries the main results. we found that the order parameters of converged states at all fillings are generally characterized by several broken symmetries: the spin-polarized (SP) state with $\left<s_3\right> \neq 0$, the valley-polarized (VP) state with $\left<\tau_3\right>\neq 0$, the intervalley coherent (IVC) state with \mbox{$\left<\tau_{1,2}\right>\neq 0$}, the Kramers intervalley coherent (KIVC) state with \mbox{$\left<\tau_{1,2}\otimes\Sigma_3\right>\neq 0$}, interorbital intervalley coherent (inter-orbital IVC) state with $\left<\tau_{2}\otimes\Sigma_2\right>\neq 0$, and the orbital-polarized (OP) state with $\left<\Sigma_3\right>\neq 0$. Depending on the filling, we found both the correlated metallic states and insulators with broken symmetries. Certain semimetallic states near charge neutrality preserve the $C_{2z}\trs$ symmetry, while all other orders break the $C_{2z}\trs$ symmetry.

\subsection{Hartree--Fock results at all integer fillings}

\begin{figure*}[tb!]
    \centering
    \includegraphics[width=0.8\linewidth]{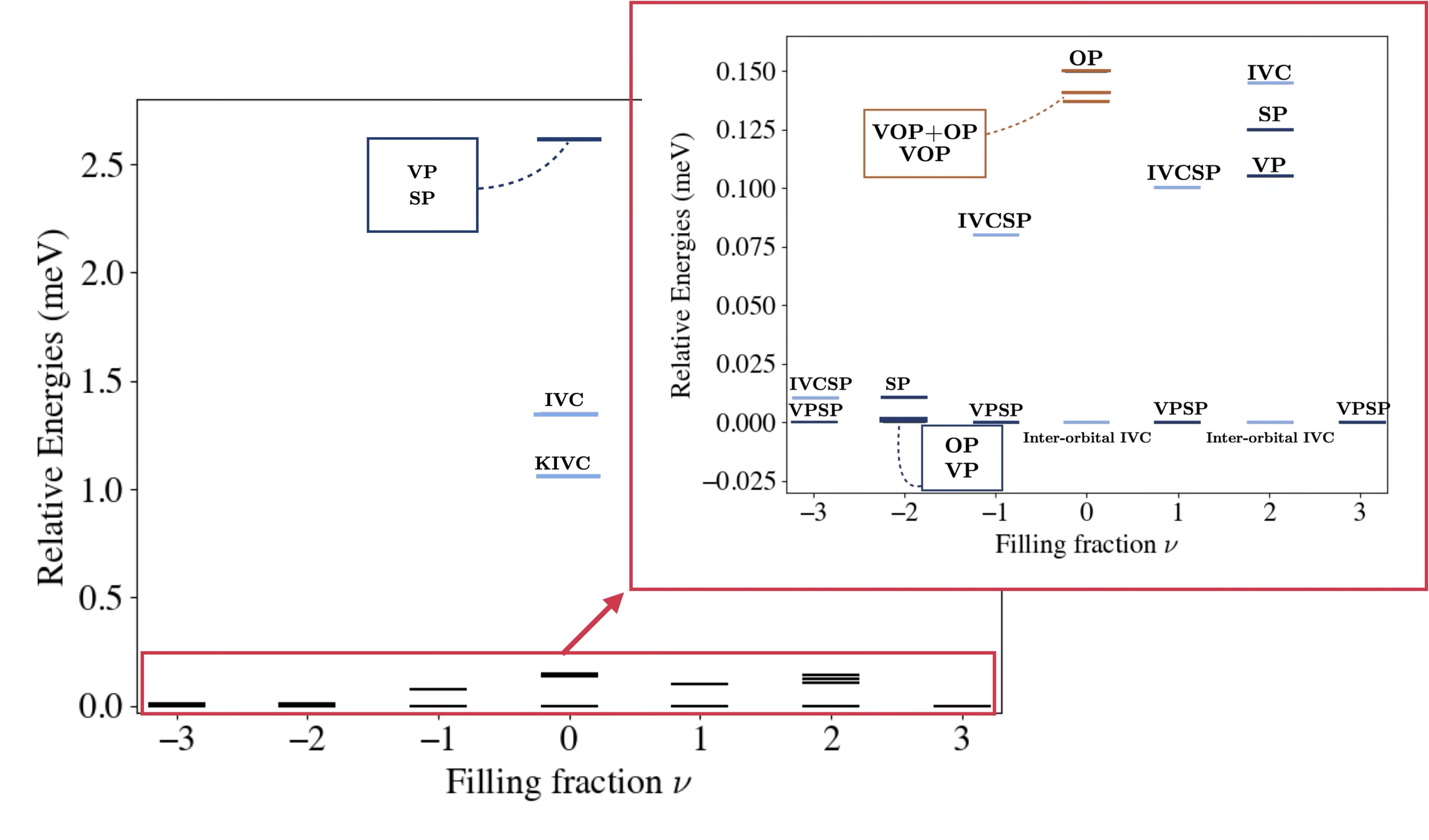}
    \caption{Relative energies of converged ordered states. and schematic phase diagram of $\theta = 1.10^{\degree}$ 8-orbital model. The inset shows details of relative energies of competing orders.}
    \label{fig:summary}
\end{figure*}

\begin{figure*}[!htb]
    \includegraphics[width=0.8\linewidth]{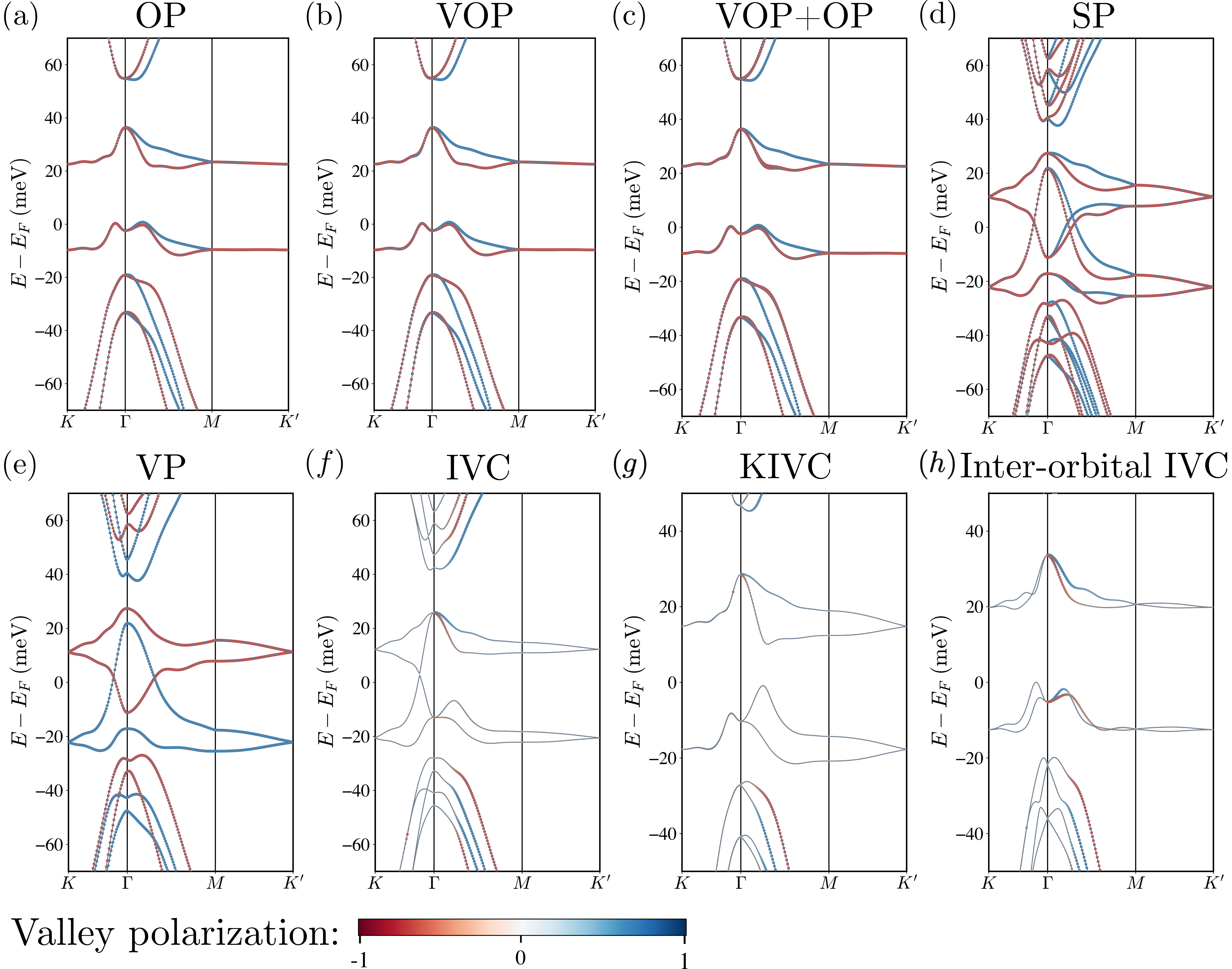}
    \caption{Ground state candidates at charge neutrality for TBG 8-orbital model. The color bar shows the valley polarization strength for each k point. (a) Orbital polarized state (OP). (b) Valley-orbital polarized state (VOP). (c) Valley-orbital polarized and orbital polarized state (VOP+OP).  (d)Spin polarized metal (SP). (e) Valley polarized semimetal (VP). (f) Inter-valley coherent state (IVC). (g) Kramers inter-valley coherent state (KIVC). (h) inter-orbital inter-valley coherent state (inter-orbital IVC).}
    \label{fig:charge_neutral}
\end{figure*}

By performing Hartree--Fock calculations on the interacting 8-orbital model in Eq.~\eqref{eq:Hfull}, we analyzed the solutions at various integer fillings. Our results are summarized in Table~\ref{table:iii} and in Fig~\ref{fig:summary}. In some cases, we were able to find a well-defined ground-state that spontaneously breaks the symmetries of the model; while in other cases, we identified several candidate states nearly degenerate in energy. The resulting symmetry-broken states and the characteristic energy differences are outlined in Fig~\ref{fig:summary}. At certain fillings, in particular $\nu=-3, -2$, we found several candidate states that have nearly identical total energies, so the competitions of these symmetry-broken states are possible.
we found that both insulating and (semi)metallic ground states are stabilized, depending on the filling fraction of the central bands. QAH solutions are also found. Below, we analyse in detail our results for all integer fillings from $\nu = -3$ to $\nu = +3$.\\

\textbf{Numerical results at filling $\boldsymbol{\nu=0}$}. At the charge neutral point, our calculations found insulating states and semimetallic states. 
We generated several initial guesses by the symmetry argument as discussed in Sec.~\ref{sec:intro}, and put them into the HF self-consistent cycles. Eight different converged ordered states are obtained, whose band structures are shown in Fig.~\ref{fig:charge_neutral} -- the orbital polarized (OP) state, the valley-orbital polarized (VOP) state, a polarized state in which the up- [down-] spin sector is orbital [valley-orbital] polarized (VOP+OP state), the spin-polarized (SP) semimetallic state, the valley polarized (VP) semimetallic state, the inter-valley coherent (IVC) state, the Kramers inter-valley coherent (KIVC) state, and inter-orbital inter-valley coherent state.

The first three ordered states OP, VOP, and VOP+OP shown in Fig.~\ref{fig:charge_neutral} exhibit similar band structures, with nearly degenerate total energies (energy differences within 0.13~meV, as shown in Fig.~\ref{fig:summary}). Among them, VOP has the lowest total energy. However, according to Table \ref{table:iii}, these three states have different total Chern numbers: $C=0, 4, 2$ respectively.

The semimetallic states in general have higher total energy compared with gapped states. One would expect opening a gap when the system is fully spin-polarized or valley-polarized due to Hund's rule, but this is not the case for our VP and SP states at charge neutrality. While it is true that the exchange interaction is repulsive between two valleys and two spins species, its magnitude is however too small to open up the gap at the $\Gamma$ point. 
We note that the semimetallic VP and IVC symmetry-broken solutions preserve the $C_{2z}\trs$ symmetry, resulting in Dirac points at $K$ points. SP state has the similar structure. It breaks the $C_{2z}\trs$ symmetry due to spin polarization. However, if only choosing one spin flavour, and making $\trs^2 = 1$, the spinless band has the $C_{2z}\trs$ symmetry.

The last two states, KIVC and inter-orbital IVC states, are both insulators. The inter-orbital IVC state has the lowest total energy, while the KIVC state has a total energy that is 1~meV larger than that of the inter-orbital IVC state. Neither state possesses Chern numbers. Under time-reversal symmetry operations, both the IVC and inter-orbital IVC states remain invariant, while the KIVC state does not respect this symmetry. The wave functions of the IVC states can be examined in Appendix \ref{app:ivcwf}.

Therefore, our results suggest that orbital orders like $\tau_{1,2}\otimes \Sigma_3$ (KIVC), $\tau_{2}\otimes \Sigma_2$ (inter-orbital IVC), $\tau_3\otimes \Sigma_3$ (VOP) and $\Sigma_3$ (OP) are the reason for generating insulating states. The mechanism is explained as follows. The interaction-driven symmetry breaking in orbital-polarized channels can be understood through a  reduced Hamiltonian composed of $p_{\pm}$ orbitals only, which is obtained by truncating the 8-orbital single-particle Hamiltonian. 
In the non-interacting limit,  such an \textit{effective} Hamiltonian  has the general form, 
\begin{align}
H_{\text{eff}}(\bm k) = \sum_{\alpha, \beta = 0}^3 h^{\alpha \beta}(\bm k)    
\end{align}
where 
\begin{equation}
h^{\alpha \beta}(\bm{k})= \sum_{s = \uparrow, \downarrow}  t^{\alpha \beta}(\bm{k})\psi_s^{\dagger}(\bm{k})~\Gamma^{\alpha \beta}~ \psi_s(\bm{k}),
\end{equation}
with $\Gamma^{\alpha \beta}=\tau_\alpha \otimes \Sigma_{\beta}$ are the generators of $SU(4)$, $\psi_s= (c_{1, +, s} \quad c_{2, +, s} \quad c_{1, -, s} \quad c_{2, -, s})^\intercal$, and $t^{\alpha \beta}(\bm{k})$ encodes the respective hopping amplitudes. 
We note that, here, the spin degrees of freedom are trivially summed over because the Hamiltonian is identity in spin space, as the spin-orbit coupling is negligible in graphene. 
For the model we study, the only non-vanishing terms in $H_{\text{eff}}$ correspond to $(\alpha, \beta) = (0,0), (0, 1), (0,2), (3,1), (3,2)$, and $(3,0)$.  
While the first five terms obtain contributions from nearest and next-nearest neighbor hoppings, $t^{30}$ lacks contributions from nearest neighbor hoppings.
Therefore, $t^{30}$ is exponentially suppressed compared to the remaining hopping amplitudes in $H_{\text{eff}}$.
We restrict the following discussion to nearest-neighbor hoppings only, such that 
\begin{align}
H_{\text{eff}} \to  H_{\text{eff}, \text{NN}} = h^{00} + h^{01} + h^{02}+ h^{31}+ h^{32},
\label{eq:Heff}
\end{align}
where we have suppressed the $\bm k$-dependence for notational simplicity.
We will comment on the potential impact of further neighbor hoppings at the end of the present analysis.
We note that the four bands produced by diagonalizing $H_{\text{eff}, \text{NN}}$ are twofold spin-degenerate.

The term $t^{00}(\bm{k})$, while being of the same order as the other four hopping amplitudes,  does not directly control the gapping of the Dirac points in the mini-Brillouin zone.
This is because it effectively serves as a local chemical potential inside each unit-cell, and commutes with all patterns of internal symmetry breaking.
Moreover, in the vicinity of the $K$ points of the mini-Brillouin zone, $t^{00}(\bm{k})$ is only weakly $\bm k$-dependent.

If the order parameter of a symmetry-broken state globally anti-commutes with $H_{\text{eff}}$, then it follows from the properties of the Clifford algebra that the ordered state must be gapped. 

For our effective model in Eq.~\eqref{eq:Heff}, we found the following (anti-)commutation relations:
\begin{align}
& \left \{ (H_{\text{eff}, \text{NN}} - h^{00}), \Gamma^{03} \right\} = 0, 
    \nonumber\\ %\quad
& \left \{ (H_{\text{eff}, \text{NN}} - h^{00}),\Gamma^{33} \right\} = 0, \nonumber \\
& \left [ (H_{\text{eff}, \text{NN}} - h^{00}),\Gamma^{30} \right] = 0.
\end{align}
 The generators $\Gamma^{03}$ and $\Gamma^{33} $ correspond to the OP and VOP order parameters, respectively, and the anticommutation agrees with our finding of the gap opening at charge neutrality. The semimetallic states VP (generated by $\Gamma^{30}$) and SP (generated by $s_3$ Pauli matrix)  have order parameters that commute with the Hamiltonian, thus remaining gapless.

The size of the gap opened by the OP and VOP order parameters is controlled by the strength of $H_C$, $U_{11} \approx U_{22}$ which is much larger than $\text{max}\left[t^{30}(\bm k)\right]$.
Therefore, switching-on of the next-nearest neighbor hoppings acts as weak perturbations, and does not qualitatively change the above conclusions. 

\vspace{4mm}
\textbf{Numerical results at filling $\boldsymbol{\nu=\pm2}$}. 
For the filling $\nu =+2$, the VP insulator, SP insulator, inter-orbital IVC insulator, and IVC+VP metallic solutions are found (see Fig.~\ref{fig:p2}a,b,c,d), with small energy differences (around 0.5~meV) among them. The three insulating states have small band gaps around 3~meV.
Among the four orders, the inter-orbital IVC state turns out to have the lowest total energy with $|C| = 0$. And VP state realizes an anomalous quantum Hall state, with the Chern number $|C|=2$ (by contrast, the SP insulating state is topologically trivial with $C=0$).

At the filling $\nu=-2$, VP, SP, and OP ordered metallic states are obtained. Among the three states, the VP state has the lowest total energy, however  all three states have  very small energy differences within about 0.01~meV. 
As can be seen 
from the band structures displayed in Appendix~\ref{app: bandfig} (Fig.~\ref{fig:p2}e,f,g), all three states are semimetallic -- they are not fully gapped between the hole band centered at the $\Gamma$ point and the electron band minimum centered on the $\Gamma-M$ line.

This phenomenon shows that higher-lying (non-flat) bands participate in the formation of the order, which is the consequence of small band gaps around the central bands of our model, and the fragile topology of the system. Compared with the BM models, the 8-orbital model is away from the chiral limit and has smaller gaps. 

Although at hole doping $\nu = -2$ all the states we obtained are metallic, but it is still possible to manually separate the higher lying conduction bands and calculate the Chern number of the valence bands as a reference. Defined in this fashion, the VP and SP states have $|C| = 2$ and $C = 0$ respectively, same as the Chern numbers at $\nu = +2$ filling. We thus conclude that the VP semimetallic state is a `failed' Chern insulator, with non-trivial (non-quantized) anomalous Hall response.

Unlike the approximate BM model, which is particle-hole symmetric, the more realistic 8-orbital model from Carr \emph{et al}~\cite{carr2019derivation} is not. Thus generically, one does not expect to find the same solution at filling $\pm |\nu|$. This is manifestly the case in our calculations at $|\nu| = 2$, where we found the candidate ground states to all be  metallic on the hole-doped $\nu=-2$ side, in contrast to the aforementioned opening of the gap at $\nu=2$.

\vspace{4mm}
\textbf{Numerical results at filling $\boldsymbol{\nu=\pm1}$}. At fillings $\nu = \pm1$, we obtained two states on both the electron and hole-doped sides as shown in Appendix~\ref{app: bandfig} (Fig.~\ref{fig:=-1}): the intervalley-coherent spin-polarized (IVCSP) state and a valley- and spin-polarized (VPSP) state. However, only one insulating state is found -- the VPSP state at $\nu=-1$ with a direct gap around 5.6~meV. On the other hand, the IVCSP state (whose energy is 2~meV higher) has a biquadratic band touching at the $\Gamma$ point where the gap closes.
By contrast, the $\nu=+1$ states are metallic states. 
Finally, we noted that all the aforementioned states  break the $C_{2z}\trs$ symmetry. The insulating VPSP state at $\nu = -1$ is a QAH state with Chern number $|C| = 1$.

Note that the particle-hole asymmetry characteristic of the non-interacting 8-orbital model remains clearly evident in the symmetry-broken states. This underlies our finding that the resulting Hartree--Fock band structures in the presence of the interactions appear very asymmetric between $\nu=-1$ (Fig.~\ref{fig:=-1}a,b) and $\nu=+1$ (Fig.~\ref{fig:=-1}c,d), similar to what we found for $|\nu|=2$ above.

\begin{figure}[t]
    \centering
    \includegraphics[width=\linewidth]{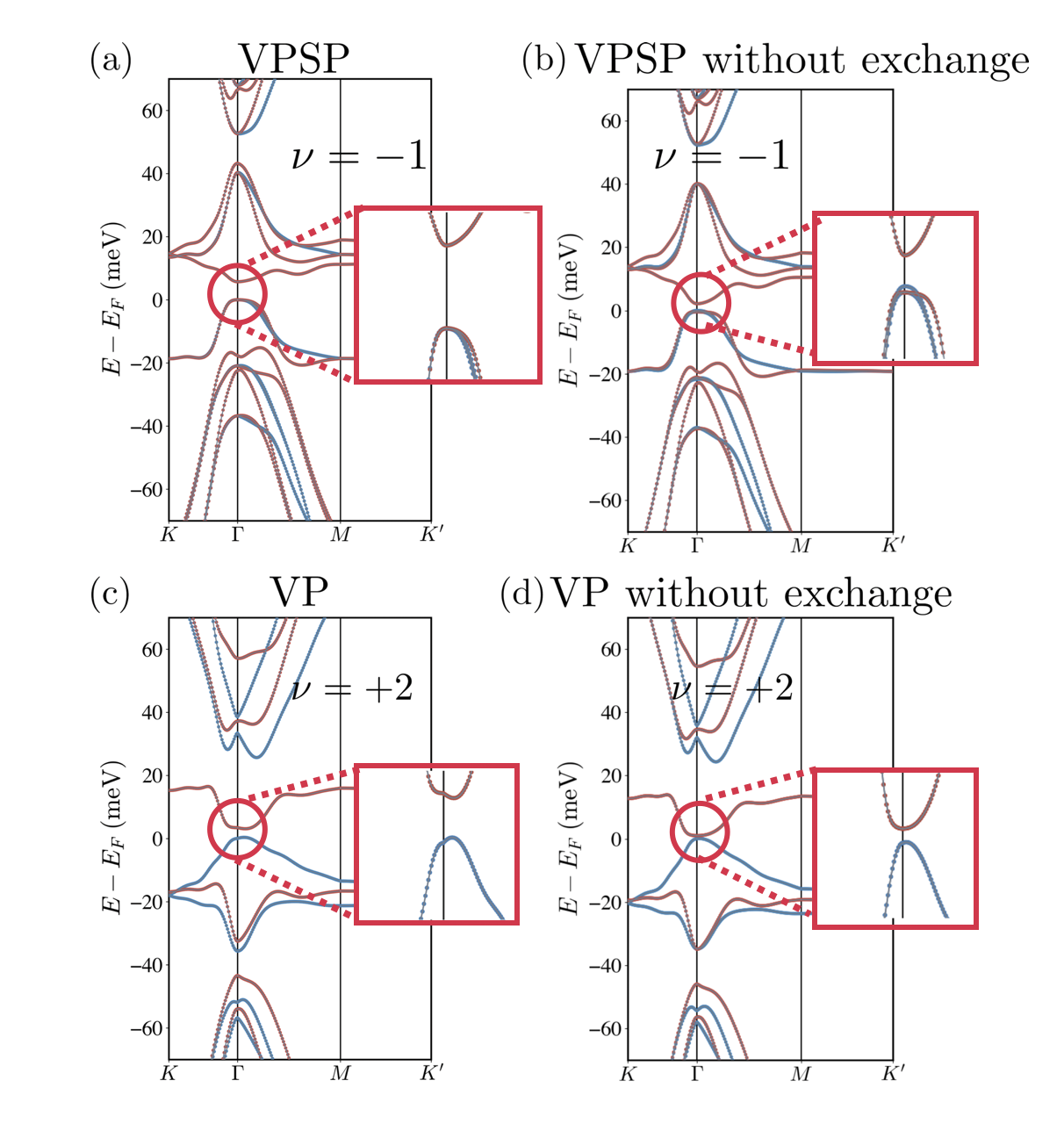}
    \caption{Comparison of insulating band structures with and without exchange interaction. (a),(b) Valley polarized and spin polarized state at $\nu =-1$. The exchange interaction enlarges the band gap. (c),(d) the valley polarized state at $\nu = +2$. The exchange interaction opens a gap around fermi level.}
    \label{fig:ex_and_withoutex}
\end{figure}

\vspace{4mm}
\textbf{Numerical results at filling $\boldsymbol{\nu=\pm3}$}. Fewer states are found at filling $\nu=\pm3$. As shown in Fig.~\ref{fig:+-3}, only a VPSP ordered state is found at filling $\nu=+3$. The state is not fully gapped, but if we separate the bands and calculate the total Chern number of the valence bands, we obtained $|C| = 1$, illustrating that this is an example of a `failed' Chern insulator similar to the VP state we found at $\nu=-2$.

By contrast, at filling $\nu = -3$, two metallic states IVCSP and VPSP are obtained. The two states show the competing feature as their total energy difference is within 0.01~meV.

\subsection{The significance of Fock terms and exchange interactions}

The previous work~\cite{Calder2020Interactions} studied the magic-angle TBG system without Fock terms and exchange interactions. The resulting  state found in the self-consistent Hartree approximation is always a metallic state for all fillings, which cannot explain most of the experimental results. 

Fock terms are important for two reasons. On the one hand, Fock terms tend to create symmetry-broken states that are comparable to experimental results. On the other hand, keeping Fock terms makes the HF state a variational Slater determinant and obey Wick's theorem, which means systematic improvements beyond HF can be made in the future based on HF states found in this work.

We emphasize that in the multi-orbital setting such as realized in TBG, the ``exchange interaction'' is not synonymous with the ``Fock term'' that is occasionally used, somewhat confusingly, to describe the exchange. Indeed, the distinction between the direct Coulomb interaction $H_C$ in Eq.~\eqref{eq:direct} and the exchange interaction $H_X$ in Eq.~\eqref{eqn:X} is made at the level of the Hamiltonian, regardless of the approximate method such as Hartree or Hartree--Fock used to optimize the ground state. Below, we analyze the effect of the exchange interactions $H_X$ on the physics of TBG.

Given the much larger value of direct density-density interactions $H_C$ (as large as 35~meV, see Fig.~\ref{fig:sg_Coulomb}) compared to the strength of exchange interactions $H_X$ ($\lesssim 3$~meV, Fig.~\ref{fig:sg_exchange}), some of the previous studies~\cite{Calder2020Interactions} have neglected the exchange interactions for this reason, performing calculations with only density-density interactions. 
It is reasonable to ask whether exchange interactions play a role in our HF calculations, if any, in stabilizing the various symmetry-broken states we had identified. 

To answer this question, we analyzed the results of our modeling with only the direct density-density interactions in Eq.~\eqref{eq:direct} vs. the full interacting Hamiltonian including the exchange terms $H_X$ as in Eq.~\eqref{eqn:X}.
We found that the exchange interaction is essential for obtaining the correlated insulating states we reported, except perhaps for the cases at charge neutrality. For example, comparing the two approaches at filling $\nu = +2$ in Fig.~\ref{fig:ex_and_withoutex}(c)(d), 
we found that excluding the exchange interaction would result in the gap closing at the $\Gamma$ point. 
The same behavior is found at filling $\nu = -1$ as shown in Fig.~\ref{fig:ex_and_withoutex}(a) and (b), where the band gap at $\Gamma$ point is much larger if the exchange interactions are included.

These numerical observations can be simply explained by the generalized Hund's interaction involving the valley and spin degrees of freedom -- the corresponding indices are treated on an equal footing in the exchange Hamiltonian in Eq.~\eqref{eqn:X}. 
The Hund's effect originates from the terms that have equal indices $\tau = \tau^\prime$ and $s = s^\prime$. In other words, the fully spin-polarized or valley-polarized bands are preferentially filled to lower the total energy. Of course the $\mathbf{k}$ dependence of the kinetic energy makes the picture more complicated, but the above argument still applies qualitatively. 
By the same token, the Hund's rule would suggest that intervalley-coherent states are less energetically favorable, which is indeed reflected in our numerical findings summarized in Fig.~\ref{fig:summary}.

\section{COMPARISON WITH OTHER HF CALCULATIONS}\label{ref:HFComparision}
The eight-orbital model incorporates more Wannier functions per spin and valley, making it a UV theory of the THF model. It also contains additional information inherited from \textit{ab initio} calculations such as lattice relaxation and graphene $sp^2$ orbital potential.

From the existing literature, we aim to compare our HF results with those of the BM model and the THF model. The early work \cite{bultinck_ground_2020} numerically showed that the particle-hole symmetric BM model has KIVC state as the lowest energy state. Subsequent numerical calculations \cite{liu2020nematic} showed that a semi-metallic state can be a ground state candidate at charge neutrality. Our results demonstrate that several ground state candidates at charge neutrality are indeed very close to each other within the energy window of 0.1~meV (see Fig.~\ref{fig:summary}). The KIVC state is also obtained, but this state has relatively higher total energy than the proposed inter-orbital IVC state that we identified as the ground state. The semi-metallic states such as IVC, VP, and SP states in our results all have higher energies exceeding 1~meV.

For all fillings, the comparison with other typical HF calculations using the BM and THF models is listed in Table \ref{tab:comparision}. We observe that all calculations are sensitive to model parameters, such as the presence of the hBN substrate, lattice strain, and lattice relaxation. At finite temperatures, we expect fluctuations on top of the mean-field order parameters. Additionally, the two THF models do not yield identical results, indicating that the model parameters significantly affect HF calculations, despite the similarity in the methods used to obtain these models. Our results suggest that the asymmetric particle-hole feature of our \textit{ab initio} derived model could lead to a novel time-reversal invariant inter-orbital IVC state at charge neutrality. At finite fillings, the hole and particle doping phases are similarly distinct, as observed in experiments, which will be shown in the next section. Importantly, while particle-hole symmetric models have insulating phases at integer fillings, our calculations show that the particle-hole asymmetric band can lead to correlated metallic states, as summarized in Fig.~\ref{fig:summary} and surrounding text.

\begin{center}
    \begin{table*}
        \centering
        \begin{tabular}{c|c|c|c|c}
        \hline
        \hline
             $\nu$& $\pm 3$ & $\pm 2$ &$\pm 1$ & 0\\
             \hline
             This work \vtop{\hbox{\strut $+$} \hbox{\strut$-$ }}& \vtop{\hbox{\strut VPSP metal} \hbox{\strut VPSP metal}} & \vtop{\hbox{\strut inter-orbital IVC insulator} \hbox{\strut \quad \quad \quad VP metal}}&  \vtop{\hbox{\strut VPSP metal} \hbox{\strut QAH insulator}} & \vtop{\hbox{\strut inter-orbital IVC insulator} \hbox{\strut inter-orbital IVC insulator}}\\
              \hline
              
             Ref.~\cite{zhang_HF_2020} (insulators)&$C_{2z}\mathcal{T}$-broken insulator& KIVC&KIVC&KIVC\\
           
             \hline
             Ref.~\cite{wagner2022global,kwan2021kekule} (insulators) &QAH,IKS& KIVC,IKS&IVC+QAH,IKS&KIVC, SM\\
            \hline
             Ref.~\cite{song2022magic} (insulators) &VP&KIVC&VP + KIVC&KIVC\\
            \hline
             Ref.~\cite{Shi_heavy_fermion2022} (insulators) &KIVC&KIVC&KIVC&KIVC\\
             
             \hline
             \hline
        \end{tabular}
        \caption{Comparison of HF ground state candidates with typical theoretical researches. QAH,IKS and SM respectively represent quantum anomalous Hall, incommensurate Kekul\'e spiral, and semi-metal. In Ref.~\cite{wagner2022global,kwan2021kekule}, the ground states depend on lattice strain. With energy differences among other candidate states falling within 0.2~meV,
including our own results, it is expected that small temperature changes
will give rise to fluctuations of orders.}
        \label{tab:comparision}
    \end{table*}
\end{center}

\section{COMPARISON WITH EXPERIMENTS}\label{sec:comparison}

The particle-hole asymmetric nature of the correlated states found in our calculations broadly agrees with the absence of particle-hole symmetry in the experiments. In this section, we provide a detailed comparison with experimental data, highlighting the agreement with our computational results.

We begin with a comparison at the charge-neutral point, where scanning tunneling microscopy (STM) experiments reported strong local correlations~\cite{jiang_charge_2019, xie2019spectroscopic, choi_imaging_2019, wong_cascade_2020}, while transport measurements identified semi-metallic features~\cite{cao_correlated_2018, cao_unconventional_2018, yankowitz2019tuning, zondiner_cascade_2020}. The strong correlation is characterized by the splitting of degenerate flat bands. While some experiments originally stated that this splitting is due to strong repulsive Coulomb interactions—leading to Hubbard bands and a cascade of spectroscopic transitions as a function of electron filling—analytical and numerical exact diagonalization studies \cite{Biao-TBG4, TBG6} suggest that the ground state is instead very close to a Slater determinant state that can be described by HF calculations. Our results indicate several very close in energy insulating ground state candidates at charge neutrality, showing that the ground state can depend sensitively on the model parameters. Transport measurements found Dirac-like features that have symmetry-preserving characteristics. Indeed, we also find semi-metallic states at the charge-neutral point. Crucially, our semi-metallic states exhibit spontaneous symmetry breaking in various valley and orbital channels, distinguishing them from the strongly correlated symmetry-preserving states obtained from the heavy fermion model~\cite{choukondo2023, husymmetric2023, 2023arXiv230113024D}, which should align better with experimental results.

The experiments, although dependent on twist angles and sample alignment with hBN substrates, have reported quantized Hall conductance at certain integer fillings. For instance, in transport measurements~\cite{sharpe_emergent_2019, serlin_QAH_2019}, quantized anomalous Hall (QAH) states at $\nu = +3$ were observed. Notably, our calculation yielded a ``failed" Chern insulator (in a sense that the band crosses the Fermi level but is separable from other bands) with $|C| = 1$ at this filling. In this case, fine-tuning interactions and the twist angle could lead to a gapped QAH state. Conversely, on the hole side at $\nu = -3$, our findings indicate a gap-closing metallic state. Consistent with our results, no resistance peak at $\nu = -3$ has been observed in prior studies~\cite{serlin_QAH_2019, stepanov_interplay_2020}. This dichotomy between the hole- and electron-doped side can be traced to the particle-hole asymmetric nature of our model resulting from lattice relaxation and the $sp^2$ orbital potential.

In Ref.~\cite{saito2020isospin} resistivity measurements were reported at twist angles $\theta = 1.12$, which is close to $\theta = 1.1$ considered here. 
There are clear resistivity peaks at even integer fillings, which may result from either insulating or semimetallic states -- in either case a reduction of the low energy density of states close to the Fermi level will suppress conductivity.
Our numerical results also show insulating or semimetallic features at even integer fillings, which would be compatible with the observed increase in resistivity.
Further, at $\nu = +3$ ($\nu = \pm 1$) the experimental observations support  (appears to not support, respectively) an enhanced resistivity, while the situation is less clear at $\nu = -3$. 
Our results indicate an insulating state at filling $\nu = -1$, a semimetallic state at $\nu = +3$, and metallic states at $\nu = +1, -3$.
Therefore, our results would agree  with the observation at $\nu = +1, +3$ and disagree with those at $\nu = -1$.

Our results for ground state candidates at $\nu = \pm 1$ also exhibit particle-hole asymmetry. Experimental observations at $\nu = +1$ show a correlated Chern insulator with a Chern number $C = 1$~\cite{stepanov_competing_2021}, while our findings also predict a $C = 1$ QAH state at hole doping $\nu = -1$.

At $\nu = \pm 2$, experiments typically indicate correlated insulating features, although outcomes may vary between devices~\cite{stepanov_interplay_2020}. Our calculations for $\nu = 2$ yield an inter-orbital IVC insulator without a net Chern number. On the hole side, at $\nu = -2$, we obtained a metallic state, as expected due to particle-hole asymmetry.

To conclude, the spontaneous symmetry-breaking states from our Hartree--Fock calculations explain some of the asymmetric phases observed in experiments. However, due to the symmetry-breaking nature of our metallic states, HF calculations have limitations and cannot explain the strongly correlated symmetry-preserving metallic states. These strongly interacting Fermi liquid states should be addressed with more sophisticated methods, such as for instance DMFT. Our HF calculations nevertheless underscore the importance of incorporating faithful particle-hole asymmetric models (in contrast to approximate effective models proposed elsewhere) into theoretical considerations.

\section{CONCLUSION}\label{sec:conclusion}

In this work, motivated by the challenge of accurately capturing the interactions in the magic-angle TBG without sacrificing the \textit{ab initio} perspective, we have studied a multi-orbital model that circumvents the fragile topological obstruction by enlarging the active orbital space. We first employed the localized Wannier orbitals to
numerically evaluate the matrix elements of the electron repulsion, enabling us to construct a suitably extended Hubbard model including both the direct and exchange interactions. Subsequently, we incorporated the complete Hartree and Fock terms to study this model. Specifically, we show that the particle-hole anisotropy in the bandstructure of the eight-orbital model translates to an asymmetry in the phase diagram with respect to the charge neutrality point ($\nu = 0$).

By performing Hartree--Fock calculations, we have obtained a variety of states as a function of integer fillings from $-3$ to $+3$ of the nearly-flat bands. 
These symmetry-broken ordered states originate from the valley, spin, and orbital degrees of freedom. The orbital polarization, which spontaneously breaks the discrete $C_{2z}\mathcal{T}$ symmetry, appears promptly at several fillings and is particularly important, as it is missing from the alternative treatments such as those based on the Bistritzer--MacDonald model. 
Our results are not symmetric with respect to electron and hole doping, consistent with the experimental  observations on TBG near the magic angle (see previous Section~\ref{sec:comparison} for detailed comparison with experiments). This asymmetry is natural for the model derived from first principles which lacks particle-hole symmetry, in contrast to approximate models such as the BM model that have been subject of many previous studies. 

We found symmetry-broken insulating states at $\nu=0,-1,+2$. In many cases, several ground state candidates have very close total energies, less than 0.16~meV apart, suggesting that the precise nature of the ground state depends delicately on the microscopic model parameters, which in turn depend sensitively on the twist angle, encapsulation, defects, and strain effects.
Properly including exchange interactions is crucial to obtain insulating states especially at $\nu = -1,+2$.

We found the anomalous Hall insulators among insulating fillings, characterized by a non-zero Chern number of the occupied band, in which the anomalous Hall state tends to be a ground state at $\nu = -1$. Several semimetallic states  with clearly separable but overlapping in energy  conduction and valence bands can also be described as `failed' Chern insulators, with a well defined Chern number of the valence bands. These metallic states are predicted to have (non-quantized) anomalous Hall effect due to non-zero integrated Berry curvature of the occupied bands.

Hartree-Fock is of course a limited theory for interacting systems. 
In comparison to Hartree-only approximations, the Fock term tends to create broken-symmetry mean field states, of which we find a number that roughly correspond to experimental observations; however, we view it as unlikely that the Hartree-Fock ground states are sufficiently accurate to determine the ground state with high confidence. 
Recently, \cite{2023arXiv230113024D}, there have been DMFT+Hartree studies of the same model, along with the aforementioned Hartree-only \cite{Calder2020Interactions} calculations. 
Those calculations drop both the exchange $H_X$ and the Fock term from consideration, which we show are both important to stabilizing symmetry-broken states. We note that DMFT calulations based on symmetry-broken states have been performed using the THF model \cite{rai2023dynamical}. 
In future works, it would be interesting to see if a multi-site extension of DMFT can be implemented for the eight-orbital model of TBG to capture long-range correlations, as we did in the present work.
Moreover, the effects of particle-hole asymmetry, crucial to the present study, remain to be explored with DMFT.

We note in passing that recently, Shi  \emph{et al.} \cite{shi2024moireopticalphononsdancing} incorporated the phononic modes into the eight-orbital model, motivating future works to study the pairing fluctuations and translational symmetry breaking orders such as Kekul\'e-type ordering.  

\begin{acknowledgments}
We would like to express our gratitude to Fang Xie for engaging in helpful discussions. This research was supported by the U.S. Department of Energy Computational Materials Sciences (CMS) program under Award No. DE-SC0020177. R.H. and A.H.N. were also supported by the Department of Energy under the Basic Energy Sciences Award No. DE-SC0025047. A.H.N. is grateful for the hospitality of the Aspen Center for Physics,  supported by the National Science Foundation Grant No. PHY-2210452, where a portion of this work was performed.
\end{acknowledgments}
\clearpage

\appendix

\section{Symmetries in 8-orbital TBG model}\label{app: symm}
We verified the symmetries of the non-interacting Hamiltonian. While we initially described the non-interacting Hamiltonian using a periodic gauge for the sake of convenience in HF calculations, it is more straightforward to describe the symmetry transformation using a physical gauge. These two Hamiltonians are connected through a gauge transformation denoted as $U_{\text{gauge}}(\bm{k})$. Whenever we require a specific gauge choice, we can perform the transformation as follows:

\begin{equation}
U^\dagger_{\text{gauge}}(\bm{k}) H^{\text{Periodic}}_K(\bm{k}) U_{\text{gauge}}(\bm{k}) = H^{\text{Physical}}_K(\bm{k}).
\end{equation}

We discuss the symmetry operations under physical gauge. The non-interacting Hamiltonian written in momentum space is 
\begin{equation}
    H_K = \sum_{ab\tau s} t^{\tau}_{ab}(\bm{k}) c^\dagger_{a \tau s}(\bm{k})c_{ b\tau s}(\bm{k}).
\end{equation}
We simply denote the matrix of $[t^{\tau}_{ab}(\bm{k})]$ as $h^\tau(\bm{k})$. When there is no valley dependence, index $\tau$ will be omitted. In the following verification, We use the notation $D(g)$ to label the representation matrix of a symmetry operation $g$. We consider a valley-independent symmetry operation acting on a creation operator:
\begin{equation}
    g^{-1} c^{\dagger}_{a\tau s}(\bm{k}) g =\sum_b D^{*}_{ba}(g)c^{\dagger}_{b\tau s}(g\bm{k}).
\end{equation}
This gives the definition of matrix $D(g)$. Using the symmetry condition $g^{-1} H_K g = H_K$, we could derive how the matrix $h^\tau(\bm{k})$ transforms under a symmetry operation.

The above definition assumes the operation does not mix two valleys. We consider an example that does mix two valleys---the time-reversal symmetry. The time-reversal symmetry is anti-unitary and relates two valleys
\begin{equation}
    \trs^{-1}c_{a\tau s}(\bm{k}) \trs = c_{a-\tau s}(-\bm{k}).
\end{equation}
Note here we choose the spinless anti-unitary convention $\trs^2 = 1$, but if an order is related to any spin polarization, we choose back to original definition $\trs^2 = -1$.
It is easy to verify that
\begin{equation}
    h^{\tau*}(-\bm{k}) = h^{-\tau}(\bm{k}).
\end{equation}
The relationship helps to construct the whole non-interacting Hamiltonian from a single-valley Hamiltonian.

There are three important Symmetries $C_{2x}$, $C_{2z}\trs$, and $C_{3z}$ in the magic-angle TBG system. According to the lattice structure and the symmetries of Wannier orbitals, we can write down the symmetry representation matrices easily. The Wannier basis in a single valley is denoted as $\left(\ket{1,p_+}, \ket{2,p_-}, \ket{3,s}, \ket{4,p_z}, \ket{5,p_z}, \ket{6,s}, \ket{7,s}, \ket{8,s} \right)$ where the numbers label the orbital indices as in Table \ref{table1}. We use $\mathcal{K}$ to represent the complex conjugation, in case there is a anti-unitary operator. The representation matrices of these three operators can be derived as follows:
\begin{equation}
    D(C_{2z}\trs) = \begin{pmatrix}
    
0 & -1 & 0 & 0 & 0 & 0 & 0 & 0 \\
-1 & 0 & 0 & 0 & 0 & 0 & 0 & 0 \\
0 & 0 & 1 & 0 & 0 & 0 & 0 & 0 \\
0 & 0 & 0 & 0 & 1 & 0 & 0 & 0 \\
0 & 0 & 0 & 1 & 0 & 0 & 0 & 0 \\
0 & 0 & 0 & 0 & 0 & 1 & 0 & 0 \\
0 & 0 & 0 & 0 & 0 & 0 & 1 & 0 \\
0 & 0 & 0 & 0 & 0 & 0 & 0 & 1 \\

    \end{pmatrix},
\end{equation}

\begin{equation}
    D(C_{2x}) = \begin{pmatrix}
    
0 & -1 & 0 & 0 & 0 & 0 & 0 & 0 \\
-1 & 0 & 0 & 0 & 0 & 0 & 0 & 0 \\
0 & 0 & 1 & 0 & 0 & 0 & 0 & 0 \\
0 & 0 & 0 & -1 & 0 & 0 & 0 & 0 \\
0 & 0 & 0 & 0 & -1 & 0 & 0 & 0 \\
0 & 0 & 0 & 0 & 0 & 1 & 0 & 0 \\
0 & 0 & 0 & 0 & 0 & 0 & 0 & 1 \\
0 & 0 & 0 & 0 & 0 & 0 & 1 & 0 \\

    \end{pmatrix},
\end{equation}

\begin{equation}
    D(C_{3z}) = \begin{pmatrix}
    
e^{-i2\pi/3} & 0 & 0 & 0 & 0 & 0 & 0 & 0 \\
0 & e^{i2\pi/3} & 0 & 0 & 0 & 0 & 0 & 0 \\
0 & 0 & 1 & 0 & 0 & 0 & 0 & 0 \\
0 & 0 & 0 & 1 & 0 & 0 & 0 & 0 \\
0 & 0 & 0 & 0 & 1 & 0 & 0 & 0 \\
0 & 0 & 0 & 0 & 0 & 0 & 1 & 0 \\
0 & 0 & 0 & 0 & 0 & 0 & 0 & 1 \\
0 & 0 & 0 & 0 & 0 & 1 & 0 & 0 \\

    \end{pmatrix}.
\end{equation}
Applying these three matrices to the non-interacting Hamiltonian $h(\bm{k})$, the Hamiltonian must satisfy the following relationships due to symmetry constraint:
\begin{equation}
    D^{-1}(C_{2z}\trs)h^{*}(\bm{k})D(C_{2z}\trs) = h(\bm{k}),
\end{equation}
\begin{equation}
    D^{-1}(C_{2x})h(\bm{k})D(C_{2x}) = h(C_{2x}\bm{k}),
\end{equation}
and 
\begin{equation}
    D^{-1}(C_{3z})h(\bm{k})D(C_{3z}) = h(C_{3z}\bm{k}).
\end{equation}
We verified that the non-interacting Hamiltonian has the correct symmetry.

\section{Interacting Hamiltonian}\label{app: interaction}
The four-center electron repulsion integrals are evaluated numerically using Wannier functions:
\begin{widetext}
\begin{equation}
    V^{\tau_1 \tau_2 \tau_3 \tau_4}_{a_1 a_2 a_3 a_4}(\bm{R}_1,\bm{R}_2,\bm{R}_3,\bm{R}_4) = \int \mathrm{d}\bm{r}^2\mathrm{d}{\bm{r}^\prime}^2 V_{SG}(|\bm{r}- \bm{r}^\prime |)
    \sum_{XX^\prime}\mathcal{W}^{X}_{\bm{R}_1 a_1 \tau_1}(\bm{r}) \mathcal{W}^{X}_{\bm{R}_4 a_4 \tau_4}(\bm{r} ) \mathcal{W}^{X^\prime}_{\bm{R}_2 a_2 \tau_2}(\bm{r}^\prime)\mathcal{W}^{X^\prime}_{\bm{R}_3 a_3 \tau_3}(\bm{r}^\prime ),
\end{equation}
\end{widetext}
where the Wannier functions of two valleys are related by complex conjugation:
\begin{equation}
    \mathcal{W}^{X}_{\bm{R} a }(\bm{r} ):=\mathcal{W}^{X}_{\bm{R} a \tau = +}(\bm{r} ) = \mathcal{W}^{X}_{\bm{R} a \tau = -}(\bm{r})^*.
\end{equation}
The Wannier function is proportional to an overall valley-dependent modulation:
\begin{equation}
    \mathcal{W}^{X}_{\bm{R} a \tau}(\bm{r} ) \propto \sum^3_{j=1} e^{i\tau \bm{K}^{j}_\tau \cdot(\bm{r}-\bm{R}_X )}.
\end{equation}
$\bm{K}^j_\tau$ are pristine graphene valley vectors and $\bm{R}_X$ are sublattice vectors in real-space. This modulation structure suppress those integrals whose $\tau_1 \neq \tau_4$ and $\tau_2 \neq \tau_3$, as explained in the main text. We only considered two-center integrals assuming other integrals are small. We can define the direct-channel interaction integrals $U_{a_1 a_2}(\bm{R}_1 - \bm{R}_2): =  V^{\tau \tau^\prime \tau^\prime \tau}_{a_1 a_2 a_2 a_1}(\bm{R}_1,\bm{R}_2,\bm{R}_2,\bm{R}_1)$, and exchange-channel interaction integrals $X^{\tau\tau^\prime}_{a_1 a_2}(\bm{R}_1 - \bm{R}_2):=V^{\tau \tau^\prime \tau^\prime \tau}_{a_1 a_2 a_1 a_2}(\bm{R}_1,\bm{R}_2,\bm{R}_1,\bm{R}_2)$.

The exchange interaction is important to determine the phases, although it is small than the density-density interaction $U_{ab}(\bm{R}_i - \bm{R}_j)$. We denote the exchange integral of two orbitals $a$ and $b$ located at unit cells $\bm{R}_i$ and $\bm{R}_j$ as $X^{\tau\tau^\prime}_{ab}(\bm{R}_i-\bm{R}_j)$. Because it has dependence on valleys, we can write it in two parts: the intra-valley exchange integral $J_{ab}(\bm{R}_i-\bm{R}_j)$ when $\tau = \tau^\prime$ and the inter-valley exchange integral $P_{ab}(\bm{R}_i-\bm{R}_j)$ when $\tau\neq\tau^\prime$. The matrix form of the integral is written as
\begin{equation}
    \left[X^{\tau\tau^\prime}_{ab}(\bm{R}_i-\bm{R}_j)\right] = 
    \begin{pmatrix}
        J_{ab}(\bm{R}_i-\bm{R}_j) &  P_{ab}(\bm{R}_i-\bm{R}_j)\\
         P_{ab}(\bm{R}_i-\bm{R}_j)^* & J_{ab}(\bm{R}_i-\bm{R}_j)
    \end{pmatrix}.
\end{equation}
Using the condition that the Wannier functions of two valleys are connected with a complex conjugation, these two integrals are evaluated numerically using the expressions as following:

\begin{widetext}

\begin{equation}
    J_{ab}(\bm{R}_i-\bm{R}_j) = \int \mathrm{d}\bm{r}^2\mathrm{d}{\bm{r}^\prime}^2 V_{SG}(|\bm{r}- \bm{r}^\prime |)
    \sum_{XX^\prime}\mathcal{W}^{X}_{\bm{R}_i a}(\bm{r})^*\mathcal{W}^{X}_{\bm{R}_j b}(\bm{r} ) \mathcal{W}^{X^\prime}_{\bm{R}_j b}(\bm{r}^\prime)^*\mathcal{W}^{X^\prime}_{\bm{R}_i a}(\bm{r}^\prime ),
\end{equation}

\begin{equation}
    P_{ab}(\bm{R}_i-\bm{R}_j) = \int \mathrm{d}\bm{r}^2\mathrm{d}{\bm{r}^\prime}^2 V_{SG}(|\bm{r} - \bm{r}^\prime |)
    \sum_{XX^\prime}\mathcal{W}^{X}_{\bm{R_i}a}(\bm{r})^*\mathcal{W}^{X}_{\bm{R}_j b}(\bm{r} ) \mathcal{W}^{X^\prime}_{\bm{R}_i a}(\bm{r}^\prime)^*\mathcal{W}^{X^\prime}_{\bm{R}_j b}(\bm{r}^\prime ).
\end{equation}

We set the elements $X^{\tau\tau^\prime}_{aa}(\bm{0}) = 0$ to avoid double counting of interactions that have already been calculated in density-density interaction.

Combining all the interaction terms with the tight-binding model, the full interacting Hamiltonian is

\begin{equation}\label{eqn:hx}
    H_C + H_X  = H_C + \frac{1}{2}\sum_{a,b,\tau,\tau^\prime,s,s^\prime}\sum_{ij}X^{\tau\tau^\prime}_{ab}(\bm{\Delta}\bm{R}_j):c^\dagger_{a\tau s}(\bm{R}_i)c_{b\tau s}(\bm{R}_i+\bm{\Delta}\bm{R}_j)c^\dagger_{b\tau^\prime s^\prime}(\bm{R}_i+\bm{\Delta}\bm{R}_j)c_{a\tau^\prime s^\prime}(\bm{R}_i):.
\end{equation}
\end{widetext}

\section{Hartree-Fock mean-field theory}\label{app: hftheory}
In this section, we review the HF method and basic notations.

The Hartree-Fock order parameter which can be also called the one-body reduced density matrix (1rdm) is defined in the orbital Bloch basis:
\begin{equation}
P_{a\tau s,b\tau^\prime s^\prime}(\bm{k}) = \left<c^{\dagger}_{a \tau s}(\bm{k})c_{b \tau^\prime s^\prime}(\bm{k}) - \frac{1}{2}\delta_{ab}\delta_{\tau\tau^\prime}\delta_{ s s^\prime}\right>.
\end{equation}
Here the subtraction $\frac{1}{2}\delta_{ab}\delta_{\tau\tau^\prime}\delta_{s s^\prime}$ is made to counter the double counting of the interaction. The Hartree-Fock mean-field Hamiltonian $\mathcal{H}^{HF} (\bm{k})$ has dependence on 1rdm $P_{a\tau s,b\tau^\prime s^\prime}(\bm{k})$ and is derived in the Appendix~\ref{app: hf}. Solving the self-consistent eigenvalue problem of the HF Hamiltonian
\begin{equation}
    \mathcal{H}^{HF}(\bm{k}) \ket{u_{n}(\bm{k})} = \varepsilon_n(\bm{k})\ket{u_{n}(\bm{k})},
\end{equation}
we obtained the HF band dispersion $\varepsilon_n(\bm{k})$ and corresponding band wave functions $u_{n}(\bm{k})$.
The direct inversion of the iterative subspace (DIIS)~\cite{pulay1980convergence,pulay1982improved} and energy-DIIS (EDIIS)~\cite{kudin2002black} algorithms are implemented to accelerate the convergence.

There are three important symmetries in magic-angle TBG system, in-plane $120\degree$ rotational symmetry $C_{3z}$, out-of-plane $180\degree$ rotational symmetry $C_{2x}$, and $C_{2z}T$ symmetry. $C_{2z}T$ symmetry combines in-plane $180\degree$ rotation with time-reversal symmetry. The representation matrix $D(g)$ for symmetry $g$ can be found in Appendix~\ref{app: symm}. Order parameters with breaking symmetries can lead to topological non-trivial states. For example, breaking the $C_{2z}T$ symmetry is necessary for getting non-zero Chern number state;
Therefore, We define the $C_{2z}T$ symmetry-breaking order parameter to capture the $C_{2z}T$ symmetry-breaking strength. The commutator-like order parameter denotes:
\begin{equation}
    \mathcal{D}(\bm{k}) = \left|P(\bm{k})D(C_{2z}\trs)-D(C_{2z}\trs)P(\bm{k})^*\right|
\end{equation}
where $D(C_{2z}T)$ is the representation matrix of $C_{2z}T$ symmetry.

The insulating states with non-zero Chern number are identified as QAH states. The total Chern number $C$ is calculated from the determinant of the Wilson loop matrix of the occupied bands (see Appendix~\ref{app: wilson}). Many trial states are randomly generated to initialize the self-consistent calculation, but only several relevant converged results come out. 

\section{Derivation of the symmetry breaking strength}\label{app: symmbreaking}
The symmetry breaking strength of a HF converged state can be defined using a commutation-like expression:
\begin{equation}
    \left|\left<g^{-1} c^\dagger_{a\tau s}(\bm{k})c_{b\tau^\prime s^\prime}(\bm{k})g\right>-\left<c^\dagger_{a\tau s}(\bm{k})c_{b\tau^\prime s^\prime}(\bm{k})\right>\right|,
\end{equation}
where $g$ is the symmetry operation.
Using the representation matrix $D(g)$ defined in Appendix~\ref{app: symm}, we obtain
\begin{equation}
    \begin{split}
        &\left|\left<g^{-1} c^\dagger_{a\tau s}(\bm{k})c_{b\tau^\prime s^\prime}(\bm{k})g\right>-\left<c^\dagger_{a\tau s}(\bm{k})c_{b\tau^\prime s^\prime}(\bm{k})\right>\right|\\
        =&\left|D^{*}_{ca}(g)D_{db}(g)\left<c^\dagger_{c\tau s}(g\bm{k})c_{d\tau s}(g\bm{k})\right>-\left<c^\dagger_{a\tau s}(\bm{k})c_{b\tau s}(\bm{k})\right>\right|\\
        =& \left|D^\dagger(g)P(g\bm{k})D(g)-P(\bm{k})\right|
    \end{split}
\end{equation}
for unitary transformation. Multiplying $D(g)$ on the left side, the symmetry breaking strength $\mathcal{D}(\bm{k})$ can be expressed as
\begin{equation}
    \mathcal{D}(\bm{k}) = \left|P(g\bm{k})D(g)-D(g)P(\bm{k})\right| .
\end{equation}

For anti-unitary transformation, the same procedure gives
\begin{equation}
    \mathcal{D}(\bm{k}) = \left|P(g\bm{k})D(\mathcal{K}g)-D(\mathcal{K}g)P(\bm{k})^*\right|.
\end{equation}
We mainly checked the breaking of $C_{2z}\trs$ symmetry, because breaking $C_{2z}\trs$ symmetry is the necessary condition for a finite Chern number.

\section{Hartree-Fock mean-field Hamiltonian}\label{app: hf}
We used Hartree-Fock mean-field method to study the 8-band extended Hubbard model plus exchange interactions $H = H_K + H_C + H_X$. In this section, we derived the HF self-consistent equations.

\subsection{Hartree-Fock decomposition of density-density interacting Hamiltonian}
In the first part, we derive the HF self-consistent equations of the density-density interaction part. The density-density interacting Hamiltonian is expressed as 
\begin{widetext}
    \begin{equation}
    H_C = \sum_{a,b,\tau,\tau^\prime,s,s^\prime}\left(\sum_{i} \frac{1}{2}U_{ab}(\bm{0}):n_{a\tau s}(\bm{R}_i)n_{b\tau^\prime s^\prime}(\bm{R}_i):+\sum_{ij} \frac{1}{2}U_{ab}(\bm{\Delta}\bm{R}_j):n_{a\tau s}(\bm{R}_i)n_{b\tau^\prime s^\prime}(\bm{R}_i+\bm{\Delta}\bm{R}_j): \right).
\end{equation}
\end{widetext}

Assuming there is no translational symmetry breaking, we do the Fourier transform and obtain the Hamiltonian in momentum space,
\begin{equation}
    H_C = \sum_{a,b,\tau,\tau^\prime, s, s^\prime}\sum_{\bm{q}}
    \frac{1}{2N_k} \Bar{U}_{ab}(\bm{q}) \delta\rho_{a\tau  s}(\bm{q})\delta\rho_{b \tau^\prime  s^\prime}(-\bm{q}),
\end{equation}
in which we define the charge operator $\delta \rho_{a\tau s}(\bm{q})$ as
\begin{equation}
    \delta \rho_{a\tau s}(\bm{q}) = \sum_{\bm{k}} c^\dagger_{a\tau s}(\bm{k})c_{a\tau s}(\bm{q}+\bm{k})-\frac{1}{2}\delta_{\bm{q},\bm{G}}.
\end{equation}
When doing the Fourier transform, we chose periodic gauge, so $\delta\rho(\bm{q}+\bm{G}) = \delta\rho(\bm{q})$. And $\Bar{U}_{ab}(\bm{q})$ is the Fourier transformation of real-space Coulomb repulsion, which means
\begin{equation}
    \Bar{U}_{ab}(\bm{q}) = \left(V_{ab}(\bm{0})
    + \sum_{j} U_{ab}(\bm{\Delta}\bm{R}_j)e^{i\bm{\Delta}\bm{R}_j\cdot \bm{q}}
    \right).
\end{equation}

To perform the HF decomposition, we define the one-body reduced density matrix as 
\begin{equation}
P_{a\tau s,b\tau^\prime s^\prime}(\bm{k}) = \left<c^{\dagger}_{a \tau s}(\bm{k})c_{b \tau^\prime s^\prime}(\bm{k}) - \frac{1}{2}\delta_{ab}\delta_{\tau\tau^\prime}\delta_{ s s^\prime}\right>,
\end{equation}
in which the braket is evaluated using mean-field Slater determinant.
The single-body Hamiltonian from Hartree decomposition is obtained:
\begin{widetext}
\begin{equation}
    \mathcal{H}^{(H)} =\frac{1}{N_k}\sum_{\bm{k}^\prime \bm{k}}\sum_{a,b,\tau,\tau^\prime, s, s^\prime} \Bar{U}_{ab}(\bm{0})P_{a\tau s;a\tau s}(\bm{k}^\prime)\left(c^\dagger_{b\tau^\prime s^\prime}(\bm{k})c_{b\tau^\prime s^\prime}(\bm{k})-\frac{1}{2}\right),
\end{equation}
and the Hamiltonian from Fock decomposition is
\begin{equation}
\mathcal{H}^{(F)}= -\frac{1}{2N_k}\sum_{\bm{k}\bm{k}^\prime}\sum_{ab\tau\tau^\prime s  s^\prime}\Bar{U}_{ab}(\bm{k}-\bm{k}^\prime)P_{b\tau^\prime  s^\prime;a\tau s
}(\bm{k}^\prime)\left(
    c^\dagger_{a\tau  s}(\bm{k})c_{b\tau^\prime  s^\prime}(\bm{k})-\frac{1}{2}\delta_{ab}\delta_{\tau\tau^\prime}\delta_{ s s^\prime}
    \right)+h.c..
\end{equation}
\end{widetext}
\subsection{Hartree-Fock theory for exchange interaction}
Here we show the HF decomposition of the exchange interacting Hamiltonian $H_X$ as derived in Eq.~\eqref{eqn:hx}. We do the Fourier transform using periodic gauge to obtain the momentum-space exchange interaction:
\begin{widetext}
\begin{equation}
    H_{X} = \sum_{ab\tau\tau^\prime s s^\prime}\sum_{\bm{k}_1...\bm{k}_4}\sum_{j}\frac{1}{2 N_k} X^{\tau\tau^\prime}_{ab}(\bm{\Delta}\bm{R}_j) e^{i(\bm{k}_3-\bm{k}_2)\cdot \bm{\Delta}\bm{R}_j}\delta(\bm{k}_1-\bm{k}_2+\bm{k}_3-\bm{k}_4)
        :c^\dagger_{a\tau s}(\bm{k}_1)c_{b\tau s}(\bm{k}_2)c^\dagger_{b\tau^\prime s^\prime}(\bm{k}_3)c_{a\tau^\prime s^\prime}(\bm{k}_4):.
\end{equation}
Using the same prescriptions as shown in density-density interaction, the Hartree and Fock terms are derived as 
\begin{equation}
    \mathcal{H}^{(H)}_{X} = \frac{1}{2 N_k}\sum_{\bm{k}\bm{k}^\prime} \sum_{ab\tau\tau^\prime s s^\prime}\sum_{j}X^{\tau\tau^\prime}_{ab}(\bm{\Delta}\bm{R}_j) e^{i(\bm{k}^\prime-\bm{k})\cdot \bm{\Delta}\bm{R}_j}P_{a\tau s,b\tau s}(\bm{k}) \left(c^\dagger_{b\tau^\prime s^\prime}(\bm{k}^\prime)c_{a\tau^\prime s^\prime}(\bm{k}^\prime)-\frac{1}{2}\delta_{ab}\right)+h.c.,
\end{equation}
and
\begin{equation}
    \mathcal{H}^{(F)}_{X} = -\frac{1}{2 N_k} \sum_{\bm{k}\bm{k}^\prime} \sum_{ab\tau\tau^\prime s s^\prime}
    \sum_j X^{\tau\tau^\prime}_{ab}(\bm{\Delta}\bm{R}_j)P_{a\tau s,a\tau^\prime s^\prime}(\bm{k})
    \left(c^\dagger_{b\tau^\prime s^\prime}(\bm{k}^\prime)c_{b\tau s}(\bm{k}^\prime)-\frac{1}{2}\delta_{\tau\tau^\prime}\delta_{ s s^\prime}\right)+h.c..
\end{equation}
\end{widetext}

Combining with tight-binding Hamiltonian, the total Hartree-Fock mean-field Hamiltonian can be written as
\begin{equation}
    \mathcal{H}^{HF} = H_K + \mathcal{H}^{(H)} +\mathcal{H}^{(F)} + \mathcal{H}^{(H)}_X +\mathcal{H}^{(F)}_X.
\end{equation}
The mean-field Hamiltonian $\mathcal{H}^{HF}$ is self-consistently determined by its own eigenvectors. So the self-consistent mean-field states are obtained by performing self-consistent cycles. 
The total energy is evaluated by
\begin{equation}
    E_{tot} = \left<H_K+\frac{1}{2}(\mathcal{H}^{(H)}+\mathcal{H}^{(F)}+\mathcal{H}^{(H)}_X+\mathcal{H}^{(F)}_X)\right>.
\end{equation}

\section{Wilson loops and Chern numbers}\label{app: wilson}

This section explains how to calculate Wilson loop matrices. The total Chern number is calculated by counting the windings.

We define the momentum points in the first Brillouin zone $\bm{k}=\frac{1}{2\pi}(\Tilde{k}_1\bm{G}_1,\Tilde{k}_2\bm{G_2}) = (\frac{n_x}{N_x}\bm{G}_1, \frac{n_y}{N_y}\bm{G_2})$, where $N_x$ and $N_y$ are the sampling lengths on each direction. The pair $(n_x,n_y)$ labels the discretized momentum, so $(\Tilde{k}_1,\Tilde{k}_2)$ can be viewed as reduced k points.
Supposing $u_{an}(\mathbf{k})$ is the eigenvector of band $n$ in \emph{periodic} gauge, we can define the Berry connection as $\bm{A} = (A^1, A^2)$, in which $A^1_{nm}= \sum_{a} i u^*_{an}(\Tilde{k}_1,\Tilde{k}_2)\frac{\partial}{\partial_{\Tilde{k}_1}}u_{am}(\Tilde{k}_1,\Tilde{k}_2) $.

The non-Abelian Wilson loop matrix can be written as
\begin{equation}
\begin{split}
       W_{nm}(\Tilde{k}_2) &= \mathcal{P}\exp{-i\int^{2\pi}_{0}d\Tilde{k}_1 A^1_{nm}(\Tilde{k}_1,\Tilde{k}_2) } \\
       &\approx\prod^{2\pi-\Delta\Tilde{k}}_{\Tilde{k}_1=0} \sum_{a} u^{*}_{an}(\Tilde{k}_1,\Tilde{k}_2)u_{am}(\Tilde{k}_1-\Delta \Tilde{k},\Tilde{k}_2).
\end{split}
\end{equation}
In practice, to make the numerical matrix productions stable, the singular value decomposition (SVD) is applied. The total Chern number is obtained by counting the windings of $\det W_{nm}$ of occupied bands.

\section{Approximate wave functions of IVC states}\label{app:ivcwf}
Since three different IVC states are obtained, we discuss the wave functions here in detail. Assuming the spin polarization at filling $\nu = -2$, IVC, KIVC and inter-orbital IVC trial states can be approximately written as 

{\small
\begin{equation}
        \ket{\text{IVC}} = \frac{1}{2}\left(c^\dagger_{1+\uparrow} + e^{i\phi}c^\dagger_{1-\uparrow}\right)\left(c^\dagger_{2+\uparrow} + e^{i\phi}c^\dagger_{2-\uparrow}\right) \ket{\text{vac}}
\end{equation}

\begin{equation}
        \ket{\text{KIVC}} = \frac{1}{2}\left(c^\dagger_{1+\uparrow} + e^{i\phi}c^\dagger_{1-\uparrow}\right)\left(c^\dagger_{2+\uparrow} - e^{i\phi}c^\dagger_{2-\uparrow}\right) \ket{\text{vac}}
\end{equation}
\begin{equation}
        \ket{\text{inter-orbital IVC}} = \frac{1}{2}\left(c^\dagger_{1+\uparrow} + c^\dagger_{2-\uparrow}\right)\left(c^\dagger_{2+\uparrow} - c^\dagger_{1-\uparrow}\right) \ket{\text{vac}}.
\end{equation}
}
The corresponding order parameters are
\begin{align}
    O_{\text{IVC}} = (\tau_1 \cos \phi + \tau_2 \sin \phi)\otimes \Sigma_0\\
    O_{\text{KIVC}} = (\tau_1 \cos \phi + \tau_2 \sin \phi)\otimes \Sigma_3\\
    O_{\text{inter-orbital}} = \tau_2 \otimes \Sigma_2.
\end{align}
We can clearly see that the inter-orbital state mixes the $p_+$ and $p_-$ orbitals that correspond to Chern sectors in the chiral limit, and this state is therefore sometimes referred to as “inter-Chern".
Apparently, the IVC and inter-orbital/inter-Chern IVC states are invariant under time-reversal operation, while KIVC state does not respect time-reversal symmetry. All three states carry zero Chern number. At even number fillings, the corresponding three states can be constructed similarly by considering spin degrees of freedom.

\section{Details of the band structures}\label{app: bandfig}
This appendix presents band structures obtained from HF calculations for integer fillings from -3 to +3, excluding the case of charge neutrality.

\begin{figure}[!ht]
    \centering
    \includegraphics[width=0.8\linewidth]{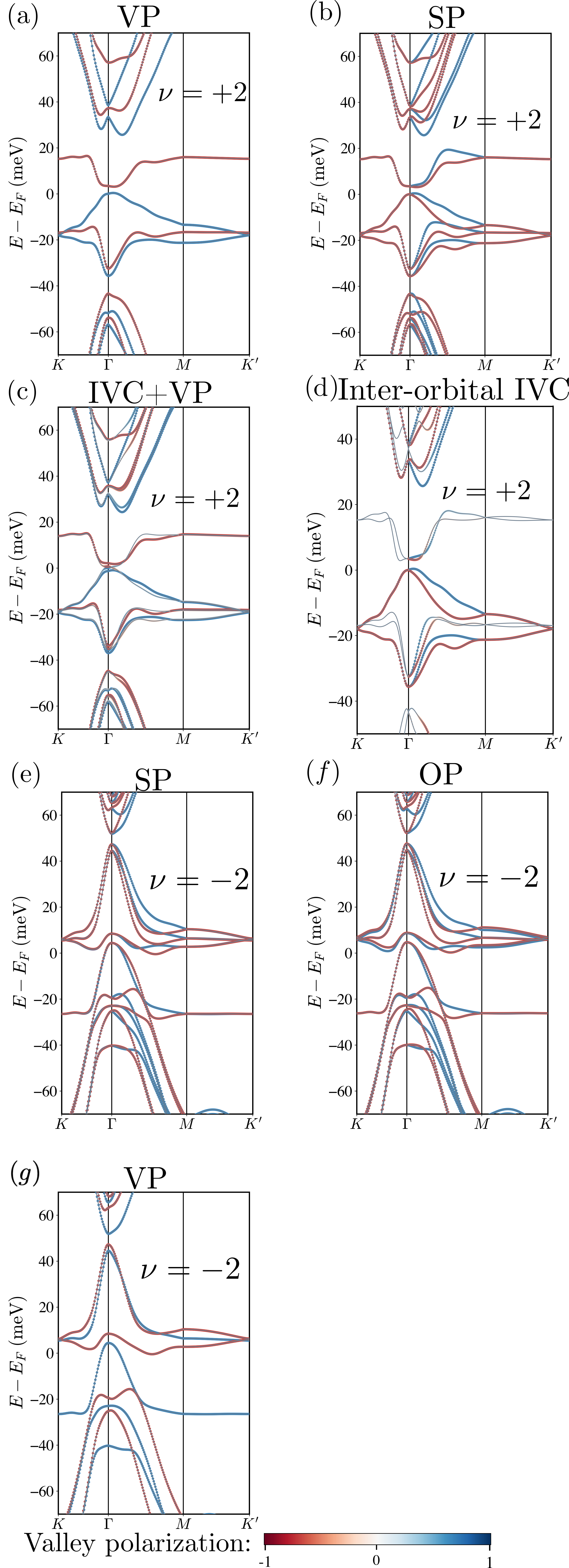}
    \caption{Band structures of self-consistent ground state candidates at $\nu=\pm 2$. The color bar shows the valley polarization
strength for each k point. (a) Valley polarized state (VP). (b) Spin polarized state (SP). (c) Inter-valley coherent and valley polarized state(IVC+VP). (d) inter-orbital IVC state. (e) Spin polarized state (SP). (f) Orbital polarized state (OP). (g) Valley polarized state (VP).}
    \label{fig:p2}
\end{figure}

\begin{figure}[!ht]
    \centering
    \includegraphics[width=\linewidth]{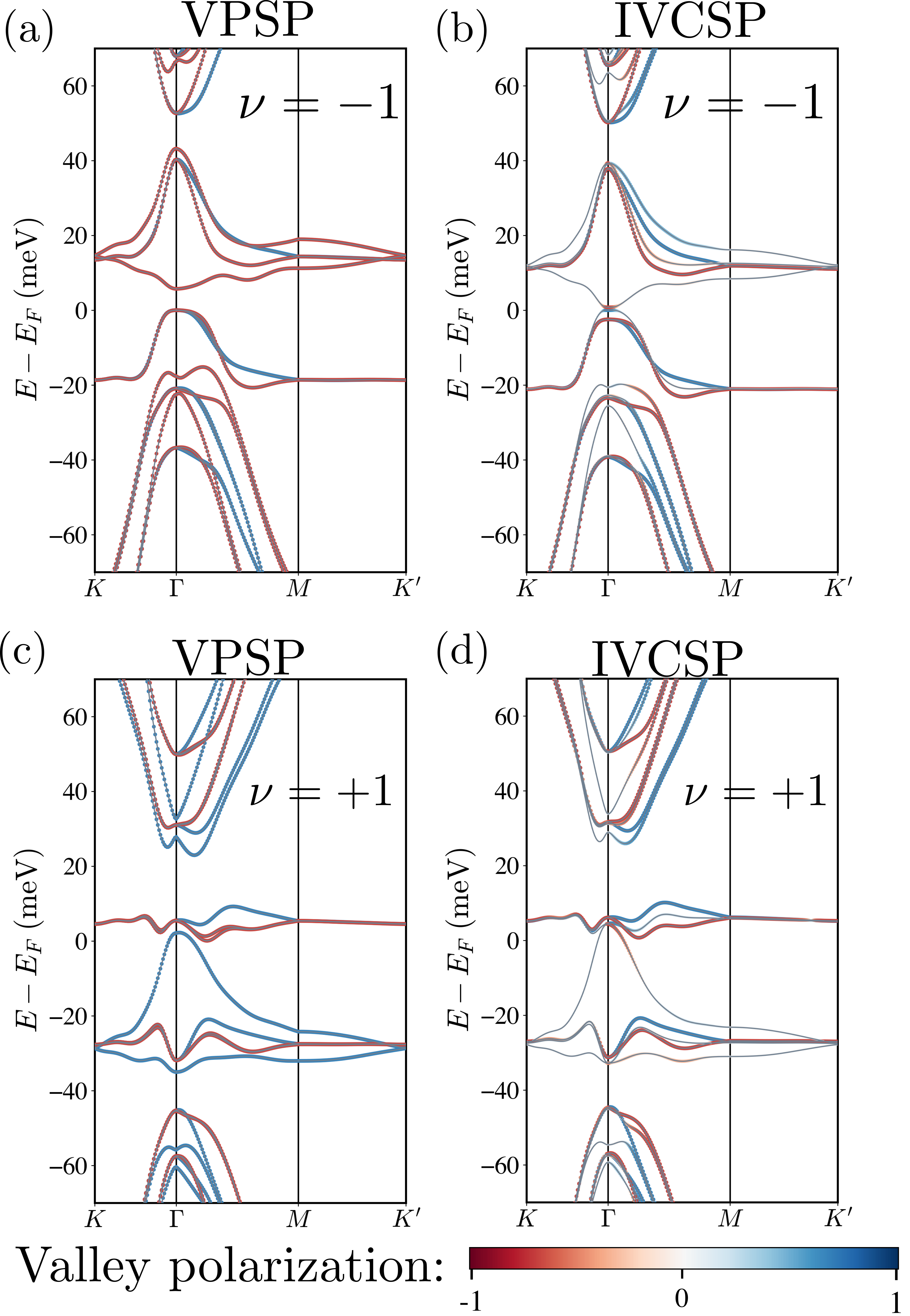}
    \caption{Band structures of self-consistent ground state candidates at $\nu=\pm 1$. The color bar shows the valley polarization
strength for each k point.  (a),(c): Valley-polarized and spin-polarized state (VPSP). (b),(d) Inter-valley coherent and spin polarized state (IVCSP). }
    \label{fig:=-1}
\end{figure}

\begin{figure}[!ht]
    \centering
    \includegraphics[width=\linewidth]{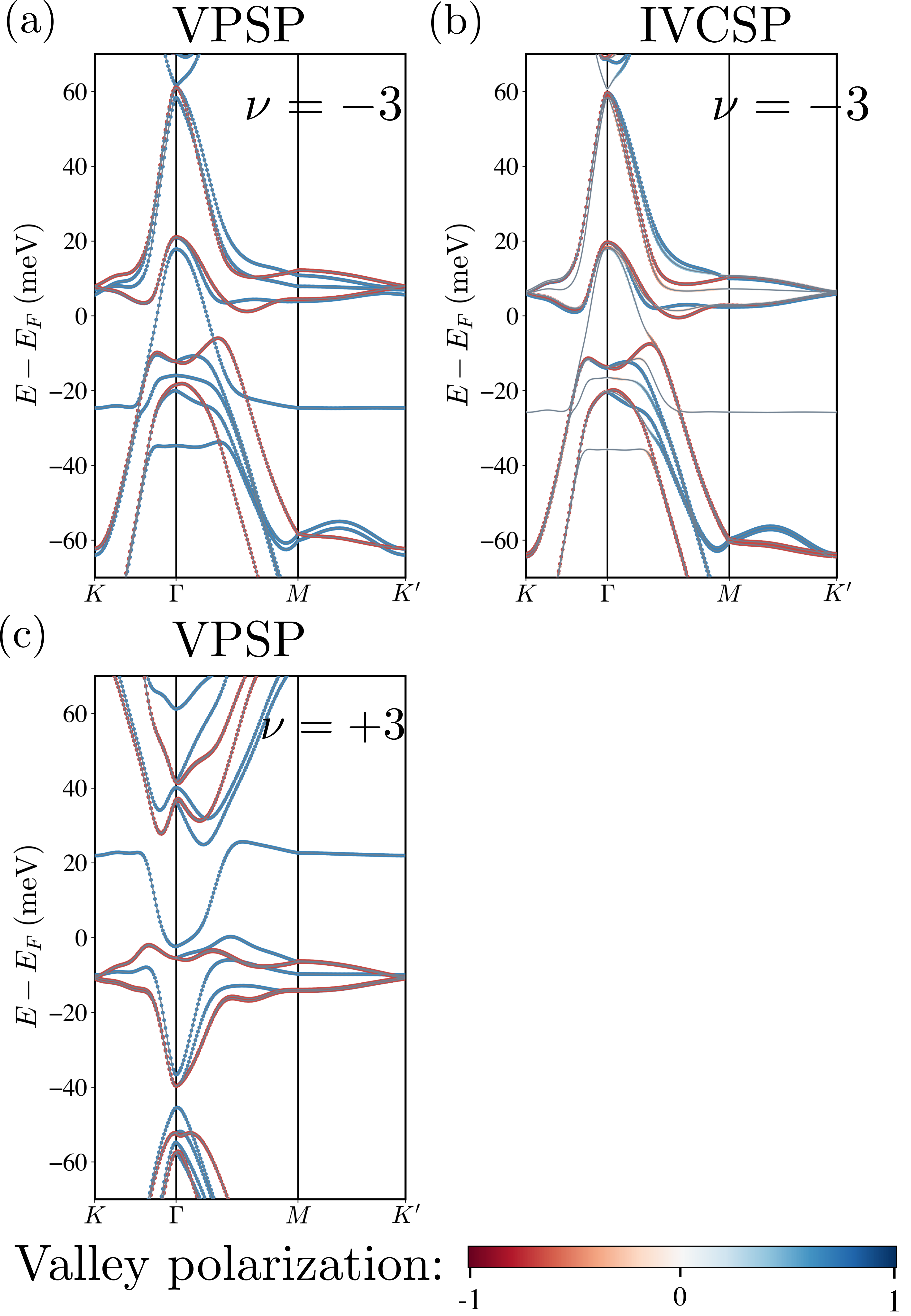}
    \caption{Band structures of self-consistent ground state candidates at $\nu=\pm 3$. The color bar shows the valley polarization
strength for each k point. (a) Valley-polarized and spin-polarized state (VPSP) for $\nu=-3$. (b) Inter-valley coherent and spin-polarized state (IVCSP) for $\nu=-3$. (c) VPSP for $\nu=+3$.}
    \label{fig:+-3}
\end{figure}

\clearpage

\bibliography{ref}
\end{document}